\documentclass[12pt, draftclsnofoot, onecolumn]{IEEEtran}
\usepackage{cite}
\usepackage{graphicx}
\usepackage{stfloats}
\usepackage{mathtools}
\usepackage{amsmath}
\usepackage{amsfonts}
\usepackage{mathrsfs}
\usepackage{amssymb}
\usepackage[caption=false,font=footnotesize]{subfig}
\usepackage{hyperref}
\newtheorem{lemma}{Lemma}
\newtheorem{corollary}{Corollary}
\newtheorem{theorem}{Theorem}
\newtheorem{remark}{Remark}
\allowdisplaybreaks
\begin{document}

\title{Decoupled Heterogeneous Networks with Millimeter Wave Small Cells}

\author{Minwei Shi, Kai Yang,~\IEEEmembership{Member,~IEEE}, Chengwen Xing,~\IEEEmembership{Member,~IEEE}, and Rongfei Fan 
\thanks{M. Shi, K. Yang, C. Xing, and R. Fan are with the School of Information and Electronics, Beijing Institute of Technology, Beijing, China, and also with Beijing Key Laboratory of Fractional Signals and Systems, Beijing, China (email: yangkai@ieee.org).}
}

\maketitle

\begin{abstract}
Deploying sub-6GHz network together with millimeter wave (mmWave) is a promising solution to simultaneously achieve sufficient coverage and high data rate. In the heterogeneous networks (HetNets), the traditional coupled access, i.e., the users are constrained to be associated with the same base station in both downlink and uplink, is no longer optimal, and the concept of downlink and uplink decoupling has recently been proposed. In this paper, we propose an analytical framework to investigate the traditional sub-6GHz HetNets integrating with mmWave small cells (SCells) with decoupled access, where both the uplink power control and mmWave interference are taken into account. Using the tools from stochastic geometry, the performance metrics of signal-to-interference-plus-noise ratio coverage probability, user-perceived rate coverage probability, and area sum rate are derived. The impact of the densification of different SCells on the network performance is also analyzed to give insights on the network design. Simulation results validate the accuracy of our analysis, and reveal that mmWave interference can not be neglected when the mmWave SCells are extremely dense and that different kinds of SCells have various effects on the network performance and thus need to be organized properly.
\end{abstract}


\IEEEpeerreviewmaketitle

\section{Introduction}
The increasing use of portable devices and multimedia applications has made an ever-growing demand for mobile data rate.
Millimeter wave~(mmWave) is a key technology to meet this challenge due to the large available bandwidth at mmWave frequencies, which would lead to higher data rate \cite{Access2013Rappaport,TWC2016Zhang}.
Although mmWave is used to be infeasible for communication due to the high near-field path loss and poor penetration through buildings, the researchers have observed that these challenges can be overcome by using highly directional antennas and beamforming \cite{rappaport2014millimeter,TC2013Hur,JSAC2017Yu_antenna}.
Another positive effect is that the mmWave interference is greatly reduced under highly directional antennas, and the mmWave networks will be noise limited rather than interference limited \cite{TWC2015Bai,TC2017mmTut_Andrews}. 

However, it is challengeable for mmWave networks to achieve high coverage probability with only mmWave base stations~(BSs) deployed \cite{TC2017mmTut_Andrews,TWC2016De_Elshaer}.
A feasible scenario is that mmWave BSs will be overlaid on traditional sub-6GHz heterogeneous networks (HetNets), where the sub-6GHz HetNets provide universal coverage and mmWave BSs provide high data rate transmission in hotspots.
The general model for HetNets is described as a combination of $K$ tiers, which are distinguished by their transmit powers, spatial densities, and propagation characters \cite{JSAC2012Femtocell_Andrews,CM2013HetNet_Andrews,TWC2012HetNet_Dhillon,TWC2012HetNet_Jo,WC2015HZhang}.

As the network goes denser and more heterogeneous, the disparity between the transmit powers of BSs is increasing, whereas the disparity between the transmit powers in uplink is roughly equal, which lead to the uplink-downlink asymmetry \cite{CM2013HetNet_Andrews}. 
Compared with the traditional coupled association strategy, which constrains the user's uplink and downlink serving BS to be the same and is no longer optimal in HetNets, the downlink-uplink decoupling~(DUDe) has emerged as an efficient approach to alleviate the uplink-downlink asymmetry and to improve the uplink performance. 
The concept of DUDe was first indicated in \cite{CM2013HetNet_Andrews}, where the author suggested to consider the downlink and uplink as two different networks, and then separately model the interference, cell association, and throughput.
Based on channel conditions, service types, and traffic loads, DUDe is able to facilitate better resource allocation among cells \cite{CM2014Green}. 
A tractable model of HetNets was proposed to characterize the network performance with decoupled access in \cite{TWC2015Singh_DUDe}, and it is shown that DUDe leads to significant improvement in rate coverage probability over the standard coupled association strategy in HetNets. 
A complete survey about DUDe can be found in \cite{CM2016DUDe}, where the authors stated that DUDe could lead to significant gains in network throughput and power consumption. 
A network consisting of hybrid sub-6GHz macrocells (MCells) and mmWave small cells (SCells) was studied in \cite{TWC2016De_Elshaer}, where the performance gains with different decoupled association strategies were investigated, and the authors observed that DUDe is a key factor in improving the uplink and downlink performance.
In DUDe networks with multi-antenna BSs, offloading the users to SCells is required in order to leverage the benefits of multiple antennas \cite{TWC2017Bacha}.

To analyze the performance of wireless networks efficiently, stochastic geometry has emerged as a unified mathematical paradigm due to its tractability and accuracy \cite{CST2017ElSawy_SGtut}. 
Specially, it was first applied to analyze the mmWave cellular networks in \cite{conf2012Akoum_mmW}, where the locations of mmWave BSs follow a two-dimensional homogeneous Poisson point process (PPP), and it is observed that mmWave could provide comparable coverage and higher data rates than microwave systems. 
Moreover, a comprehensive overview of mathematical models and analytical techniques for mmWave cellular systems was performed in \cite{TC2017mmTut_Andrews}, where the authors suggested that an mmWave network should be overlaid on a sub-6GHz network to provide high data rate in hotspots. 

Only limited work has been carried out in the hybrid sub-6GHz and mmWave cellular networks.
A HetNet consisting of sub-6GHz MCells and mmWave SCells was studied in \cite{TWC2016De_Elshaer}, where the locations of BSs are modeled as two independent PPPs, and it is observed that extremely high bias values are desirable for SCells.
A general and tractable mmWave cellular model was proposed to characterize the associated rate distribution of networks consisting of sub-6GHz MCells and mmWave SCells in \cite{JSAC2015Singh_backhaul}, and the analysis indicated that spectral efficiency of mmWave networks increases with the BS density, particularly at the cell edge.
However, sub-6GHz SCells are not taken into account in \cite{JSAC2015Singh_backhaul,TWC2016De_Elshaer}.

In this paper, using the tools from stochastic geometry, we provide a tractable framework to characterize the hybrid sub-6GHz and mmWave HetNets with decoupled access, where the user equipments~(UEs) select the downlink and uplink serving BSs separately. The main contributions of this paper are summarized as follows.
\begin{itemize}
\item
Besides the traditional MCells, both sub-6GHz and mmWave SCells are considered in our work. Moreover, the uplink power control is incorporated for uplink sub-6GHz UEs, and the power-limited  mmWave UEs transmit with constant power.
\item
Area sum rate~(ASR) is proposed to investigate the system performance of the hybrid frequency networks. The disparity of bandwidth in sub-6GHz and mmWave is considered in ASR, and it is shown that mmWave SCells are more efficiently in improving network ASR compared with sub-6GHz SCells with the same density.
\item
The general expressions for the performance metrics, including SINR coverage probability, user-perceived rate coverage probability, and ASR, are derived. 
Based on the analytical and numerical results, we investigate the impact of SCells densification, and give insights on the network design.
\end{itemize}

The rest of the paper is organized as follows. The system model is introduced in Section~\ref{sec:Model}. In Section~\ref{sec:Association}, the association probability with decoupled access is derived. The expressions of SINR coverage probability, user-perceived rate coverage probability, and ASR are given in Section~\ref{sec:Performance}. Numerical and simulation results are presented in Section~\ref{sec:Simulation}, which are followed by the conclusions in Section~\ref{sec:Conclusion}.

\section{System Model}\label{sec:Model}
We consider a two-tier sub-6GHz HetNets coexisting with mmWave SCells, where the locations of sub-6GHz MCells, sub-6GHz SCells, and mmWave SCells are modeled as homogeneous PPP $\Phi_1$, $\Phi_2$, and $\Phi_3$ with density $\lambda_1$, $\lambda_2$, and $\lambda_3$, respectively. The BSs of each tier are distinguished by their spatial densities, transmit powers, carrier frequencies as well as propagation characters. 
The UEs are spatially and independently distributed in $\mathbb R^2$ according to a homogeneous PPP $\Phi_\text{U}$ with density $\lambda_\text{U}$. The analysis is performed, without loss of generality, for a typical UE $\mathbf{y}_0$ located at the origin according to the Slivnyak-Mecke theorem \cite{chiu2013stochastic}.

It is shown that the uplink transmit power in mmWave networks is even smaller than that of sub-6GHz system \cite{AWPL2015Colombi} and power control can be neglected for mmWave networks \cite{TC2017mmTut_Andrews}. Therefore, we assume that the mmWave UEs transmit with constant power $P_u$ and that the sub-6GHz UEs utilize fractional power control~(FPC) in the uplink to partially compensate for the long-term channel variation \cite{conf2009Mullner_pc}. Given a typical UE $\mathbf{y}_0$ associated with a sub-6GHz BS in the uplink, the transmit power with FPC can be formulated as $P_u \zeta_{\mathbf{y}_0} = P_u r^{\epsilon\alpha}$, where $\zeta_{\mathbf{y}_0}$ is the FPC coefficient of the typical UE $\mathbf{y}_0$, $0 \leq\epsilon\leq 1$ is the power control fraction, $\alpha$ is the path loss exponent, and $r$ is the distance from $\mathbf{y}_0$ to its serving BS. Obviously, $\epsilon$ is equal to $0$ in the mmWave SCells.

\subsection{Directional Beamforming}
The sub-6GHz BSs are assumed to be equipped with omni-directional antennas, and the mmWave BSs are assumed to be equipped with directional antenna arrays to compensate for the high path loss. For the sake of simplicity, the directional antenna arrays are approximated by a sectored antenna model \cite{Bai2014CM_mm}, namely
\begin{align}
G_\mathrm{b}\left(\theta\right)=
	\begin{cases}
	G_\mathrm{M},&\text{if $\lvert\theta\rvert \leq {\theta_\mathrm{b}}/2$}\\ 
	G_\mathrm{m},&\text{otherwise}
	\end{cases}
\end{align}
where $\theta_\mathrm{b}$ is the beamwidth of the main lobe, and $G_\mathrm{M}$ and $G_\mathrm{m}$ are the main-lobe and side-lobe gains, respectively. When the typical UE is associated with an mmWave BS, the mmWave BS estimates the channel accurately, and then adjusts its antenna steering orientation to the typical UE to maximize the directivity gain $G_\mathrm{b}\left(\theta\right)$. The beam directions of the interference links are assumed to be independently and uniformly distributed in $\left[-\pi,\pi\right]$. Therefore, the mmWave BS's antenna gain of an interference link is $G_\mathrm{M}$ with a probability of $p_\mathrm{M} = {\theta_\mathrm{b}}/\left(2\pi\right)$, and is $G_\mathrm{m}$ with a probability of $p_\mathrm{m} = 1-{\theta_\mathrm{b}}/\left(2\pi\right)$.

\subsection{Blockage and Channel Models}
Blockage model is adopted in mmWave transmission to characterize the high near-field path loss and poor penetration through solid materials. An mmWave link can be either line-of-sight~(LoS) or non-line-of-sight~(NLoS), depending on whether the BS is visible to the UE or not. In this paper, we use 
$P_\mathrm{L}\left(r\right)$ to denote the probability that an mmWave link with length $r$ is LoS. According to the generalized blockage ball model \cite{JSAC2015Singh_backhaul}, we have
\begin{align}
    P_\mathrm{L}\left(r\right) = p_\mathrm{L} \!\cdot\! \mathbf{1}\left(r<R_\mathrm{B}\right),
\end{align}
where $\mathbf{1}\left(\cdot\right)$ is the indicator function, $R_\mathrm{B}$ is the maximum length of a LoS link, and $p_\mathrm{L}$ is the average fraction of the LoS area in the ball of radius $R_\mathrm{B}$. 
For the typical UE, mmWave BSs can be categorized into LoS BS set $\Phi_\mathrm{L}$ and NLoS BS set $\Phi_\mathrm{N}$ with distance dependent density $P_\mathrm{L}\left(r\right)\lambda_3$ and $\left(1-P_\mathrm{L}\left(r\right)\right)\lambda_3$, respectively. It is notable that $\Phi_\mathrm{L}$ and $\Phi_\mathrm{N}$ are no longer homogeneous PPP under the generalized blockage ball model.

The signals on different frequencies experience path loss with different intercepts and exponents. In mmWave networks, measurement results show a distinction between LoS and NLoS links, where the NLoS signals usually exhibit higher path loss than that of LoS signals.
Therefore, the path loss between a UE and the serving BS in the $k$th tier can be formulated as $\ell_k\left(r\right) = C_k r^{-\alpha_k}$, where $k\in\mathcal{K}=\left\{1,2,\mathrm{L,N}\right\}$, $r$ is the length of the link, and $C_k$ and $\alpha_k$ are the path loss intercept and the path loss exponent of the $k$th tier, respectively. Here, the indices of ``1'', ``2'', ``L'', and ``N'' denote the tiers of the sub-6GHz MCells, the sub-6GHz SCells, the mmWave LoS SCells, and the mmWave NLoS SCells, respectively.
The fast fading is assumed to be subject to independent and identically distributed~(i.i.d.) Rayleigh fading with unit mean, i.e., $h\thicksim \mathrm{exp}\left(1\right)$. 

\subsection{Association Strategy}
The downlink and uplink UE associations are performed based on the corresponding bias average received power~(BARP)  independently. 
Considering the typical UE, its downlink and uplink serving BSs are
\begin{align}\label{eq:BARP_DL}
	\mathbf{x}_\text{DL}^* = \mathop{\arg\max}_{\mathbf{x} \in \cup_{k\in\mathcal{K}} \Phi_k}{B_k P_{\text{DL},k} G_k \ell_k\left(\lVert \mathbf{x} \rVert\right)},
\end{align}
and
\begin{align}\label{eq:BARP_UL}
	\mathbf{x}_\text{UL}^* = \mathop{\arg\max}_{\mathbf{x} \in \cup_{k\in\mathcal{K}} \Phi_k}{B'_k P_u \lVert \mathbf{x} \rVert^{\epsilon_k \alpha_k} G_k \ell_k\left(\lVert \mathbf{x} \rVert\right)},
\end{align}
respectively,
where $B_k$ and $B'_k$ are the downlink and uplink bias values of the $k$th tier, respectively, $P_{\text{DL},k}$ and $P_u \lVert \mathbf{x} \rVert^{\epsilon_k \alpha_k}$ are the downlink transmit power of the serving BS in the $k$th tier and the uplink transmit power of the typical UE associated with BS $\mathbf{x}$ in the $k$th tier, respectively, $G_k$ is the antenna gain of BSs in the $k$th tier, and $\lVert \mathbf{x} \rVert$ is the distance from BS $\mathbf{x}$ to the typical UE.
It is worth noting that the downlink and uplink serving BSs of the typical UE may be different, i.e., $\mathbf{x}_\text{DL}^* \neq \mathbf{x}_\text{UL}^*$. With decoupled access, the uplink interference can be decreased, and thus the uplink network performance is enhanced \cite{CM2016DUDe}.

Since orthogonal multiple access is employed within a cell, intra-cell interference is mitigated here. If the typical UE is associated with the $k$th tier, the received downlink/uplink SINR can be formulated as
\begin{align}
	\text{SINR}_\text{DL} &= \frac{P_{\text{DL},k} G_k h \ell_k\left(\lVert \mathbf{x}_\text{DL}^* \rVert\right)} {\sigma_k^2 + I_{\text{DL},k}},\\
	\text{SINR}_\text{UL} &= \frac{P_u \zeta_{\mathbf{y}_0} G_k h \ell_k\left(\lVert \mathbf{x}_\text{UL}^* \rVert\right)} {\sigma_k^2 + I_{\text{UL},k}},
\end{align}
where 
$\sigma_k^2$ is the thermal noise power, and $I_{\text{DL},k}$ and $I_{\text{UL},k}$ are the downlink and uplink interference, respectively. Specifically, the downlink interference $I_{\text{DL},k}$ can be formulated as
\begin{align}\label{eq:expression_DL_interference}
	I_{\text{DL},k}\!=\!
	\begin{cases}
		\displaystyle{
			\sum_{i\in\left\{1,2\right\}} 
			\sum_{\mathbf{x}\in\Phi_i \!\backslash\! \mathbf{x}_\text{DL}^*}} 
			\!P_{\text{DL},i}  
			h_{\mathbf{x} \rightarrow \mathbf{y}_0} 
			\ell_i\! \left( \lVert \mathbf{x}\rVert \right), &\text{for $k\in\left\{ 1,2 \right\}$}\\
		\displaystyle{
			\sum_{i\in\left\{\text{L,N}\right\}} 
			\sum_{\mathbf{x}\in\Phi_i \!\backslash\! \mathbf{x}_\text{DL}^*}} 
			\!P_{\text{DL},i} 
			G_\mathrm{b}\! \left( \theta_{\mathbf{x} \rightarrow \mathbf{y}_0} \right) 
			h_{\mathbf{x} \rightarrow \mathbf{y}_0} 
			\ell_i\! \left( \lVert \mathbf{x}\rVert \right), &\text{for $k\in\left\{ \text{L,N} \right\}$}
	\end{cases}
\end{align}
where $\theta_{\mathbf{x} \rightarrow \mathbf{y}_0}$ denotes the angle between the interference link $\mathbf{x}\rightarrow\mathbf{y}_0$ and the desired link $\mathbf{x}_\text{DL}^*\rightarrow \mathbf{y}_0$. As for the uplink interference $I_{\text{UL},k}$, the applying of FPC for sub-6GHz cells makes a little difference between $k\in\left\{1,2\right\}$ and $k\in\left\{\text{L,N}\right\}$, and the expression of $I_{\text{UL},k}$ is given by
\begin{align}\label{eq:expression_UL_interference}
	I_{\text{UL},k}\!=\!\!
	\begin{cases} 
		\displaystyle{
			\!\sum_{i\in\left\{1,2\right\}} 
			\!\sum_{\mathbf{y} \in \Phi_{\!u,\!i} \!\backslash\! \mathbf{y}_0 }}
			\!P_u 
			\zeta_\mathbf{y} 
			h_{\mathbf{y} \!\rightarrow \mathbf{x}_\text{UL}^*} 
			\!\ell_i \!\left( \lVert \mathbf{y \!-\! \mathbf{x}_\text{UL}^*}\rVert \right), &\text{for $k\in\left\{ 1,2 \right\}$}\\
		\displaystyle{
			\!\sum_{i\in\left\{\text{L,N}\right\}} 
			\!\sum_{\mathbf{y} \in \Phi_{\!u,\!i} \!\backslash\! \mathbf{y}_0 }}
			\!P_u 
			G_\mathrm{b}\!\left(\! \theta_{\mathbf{y} \!\rightarrow \mathbf{x}_\text{UL}^*} \!\right) 
			\!h_{\mathbf{y} \!\rightarrow \mathbf{x}_\text{UL}^*} 
			\!\ell_i \!\left( \lVert \mathbf{y \!-\! \mathbf{x}_\text{UL}^*}\rVert \right), &\text{for $k\in\left\{ \text{L,N} \right\}$}
	\end{cases}
\end{align}
where $\Phi_{u,i}$ is the set of UEs associated with the $i$th tier, $\zeta_\mathbf{y}$ is the power control coefficient of UE $\mathbf{y}$, and $\theta_{\mathbf{y}\rightarrow\mathbf{x}_\text{UL}^*}$ denotes the angle between the interference link $\mathbf{y}\rightarrow\mathbf{x}_\text{UL}^*$ and the desired link $\mathbf{y}_0\rightarrow\mathbf{x}_\text{UL}^*$.

\section{Association Analysis}\label{sec:Association}
In order to investigate the hybrid frequency networks, we first calculate the probability of the typical UE being associated with the $k$th tier, i.e., $\mathcal{A}_{\nu,k}$, $\nu\in\left\{\text{DL,UL}\right\}$ and $k\in\mathcal{K}$. The following lemma provides the distribution of minimum distance $R_k$, which will be applied in calculating $\mathcal{A}_{\nu,k}$.

\begin{lemma}\label{lemma_mindistance}
\textit{Denote $R_k$ as the distance from the typical UE to its nearest BS in the $k$th tier, the cumulative distribution function~(CDF) of $R_k$ is given by}
\begin{align}
F_{R_k}\left(r\right) \!=\!
	\begin{cases}
	\displaystyle{1\!-\!\exp\left(-\lambda_k\pi r^2\right)}, &k\in\left\{1,2\right\}\\
	\displaystyle{1\!-\!\exp\left(-2\pi\lambda_3 \int_0^r P_k\left(x\right)x\, \mathrm{d}x\right)}, &k\in\left\{\mathrm{L,N}\right\}
	\end{cases}
\end{align}
\textit{and the probability density function~(PDF) of $R_k$ is given by}
\begin{align}
f_{R_k}\left(r\right) \!=\!
	\begin{cases}
	\displaystyle{2\pi\lambda_k r \exp\left(-\lambda_k \pi r^2\right)}, &k\in\left\{1,2\right\}\\
	\displaystyle{2\pi\lambda_3 P_k\left(r\right)r \cdot\exp\left(-2\pi\lambda_3 \int_0^r P_k\left(x\right)x\, \mathrm{d}x\right)}, &k\in\left\{\mathrm{L,N}\right\}
	\end{cases}
\end{align}
\textit{where $P_\mathrm{N}\left(r\right) = 1-P_\mathrm{L}\left(r\right)$.}
\end{lemma}
\begin{IEEEproof}
The proof can be found in \cite{TWC2012HetNet_Jo,TWC2015Bai} and  is omitted here.
\end{IEEEproof}

Denote $K_\text{DL}$ and $K_\text{UL}$ as the tier index of BSs that the typical UE is associated with in the downlink and uplink, respectively. From \eqref{eq:BARP_DL} and \eqref{eq:BARP_UL}, the event of $K_\text{DL} = k$ and $K_\text{UL} = k$ can be, respectively, described as
\begin{align}
	&\mathbf{1}\!\left(K_\text{DL}\!=\!k\right) \!=\! \mathbf{1}\!\left(\! B_k T_k R_k^{-\alpha_k} 
	\!>\! \bigcup_{ i\in\mathcal{K}\backslash k }\! B_i T_i R_i^{-\alpha_i} \!\right)\!, \\
	&\mathbf{1}\!\left(K_\text{UL}\!=\!k\right) \!=\! \mathbf{1}\!\left(\! B'_i T'_i R_i^{\left(\epsilon_k\!-\!1\right)\alpha_k} 
	\!>\! \bigcup_{ i\in\mathcal{K}\backslash k }\! B'_i T'_i R_i^{\left(\epsilon_k\!-\!1\right)\alpha_i} \!\right)\!,
\end{align}
where $T_k=P_{\text{DL},k} G_k C_k$, and $T'_k=P_u G_k C_k$. Leveraging the distance distribution $F_{R_k}\!\left(r\right)$ in Lemma~\ref{lemma_mindistance}, we can derive the association probability for each tier, as shown in Theorem \ref{theorem_association}.
\begin{theorem}\label{theorem_association}
\textit{The probability of the typical UE being associated with the $k$th tier is}
\begin{align}\label{eq:Avk}
    \mathcal{A}_{\nu,k} 
    = \int_0^\infty \prod_{ i\in\mathcal{K}\backslash k } 
    \bar{F}_{R_i}\left[\Psi_{\nu,k,i}\left(r\right)\right] 
    f_{R_k}\left(r\right)\, \mathrm{d}r,
\end{align}
\textit{where $\nu\in\left\{\text{DL,UL}\right\}$, $k\in\mathcal{K}$, $\bar{F}_{R_i}\left(r\right) = 1-F_{R_i}\left(r\right)$, and}
\begin{align}
	\Psi_{\text{DL},k,i}\left(r\right) 
	\!=\! \phi_{k,i}\left(r\right) 
	&\!=\! \left(\frac{B_i T_i}{B_k T_k}\right)^\frac{1}{\alpha_i} 
	r^\frac{\alpha_k}{\alpha_i},\label{eq:DL_Trans}
	\\
	\Psi_{\text{UL},k,i}\left(r\right) 
	\!=\! \varphi_{k,i}\left(r\right) 
	&\!=\! \left(\frac{B'_i T'_i}{B'_k T'_k}\right)^\frac{1}{\left(1-\epsilon_i\right)\alpha_i} 
	r^\frac{\left(1-\epsilon_k\right)\alpha_k}{\left(1-\epsilon_i\right)\alpha_i}.\label{eq:UL_Trans}
\end{align}
\end{theorem}
\begin{IEEEproof}
See Appendix \ref{appendix:association}.
\end{IEEEproof}

Based on the results of Theorem~\ref{theorem_association}, we can derive the distribution of the conditional distance $\mathcal{X}_{\nu,k} = \left\{ \left.R_{\nu,k}\right|{K_\nu=k} \right\}$. If the typical UE is associated with the $k$th tier in the downlink or uplink, the PDF of $\mathcal{X}_{\nu,k}$ is given by the following corollary.
\begin{corollary}\label{corollary_condistence}
\textit{The PDF of the $\mathcal{X}_{\nu,k}$ is}
\begin{align}\label{eq:corollary_PDF_fx}
	f_{\mathcal{X}_{\nu,k}}\left(x\right) = \frac{f_{R_k}\left(x\right)}{\mathcal{A}_{\nu,k}} \prod_{ i\in\mathcal{K}\backslash k } \bar{F}_{R_i}\left[\Psi_{\nu,k,i}\left(x\right)\right].
\end{align}
\end{corollary}
\begin{IEEEproof}
See Appendix \ref{appendix:condistence}.
\end{IEEEproof}

\begin{remark}\label{remark:percentage}
Obviously, the UEs can be roughly categorized into two groups, i.e., the UEs that associated to the same BS in both downlink and uplink, termed as coupled UEs, and the UEs that associated to different BSs in the downlink and uplink, termed as decoupled UEs. The percentage of the decoupled UEs is given by
\begin{align*}
	\mathcal{D} &= 1 - \sum_{k\in\mathcal{K}} \mathbb{P}\left(K_\mathrm{UL} = k, K_\mathrm{DL} = k\right)\\
	&= 1 - \sum_{k\in\mathcal{K}} \mathbb{P}\Big(
	R_k < \bigcup_{ i\in\mathcal{K}\backslash k }\varphi_{i,k}\left(R_i\right), 
	R_k < \bigcup_{ i\in\mathcal{K}\backslash k }\phi_{i,k}\left(R_i\right) \Big)\\
	&\overset{(a)}{=}\! 1 - \sum_{k\in\mathcal{K}} \prod_{ i\in\mathcal{K}\backslash k } \mathbb{P}\Big(
	R_i > \max\Big\{\varphi_{k,i}\left(R_k\right), \phi_{k,i}\left(R_k\right)\Big\}
	\Big)\\
	&\overset{(b)}{=}\! 1 \!-\! \sum_{k\in\mathcal{K}}\! \int_0^\infty \!\!\prod_{ i\in\mathcal{K}\backslash k }
	\bar{F}_{\!R_i}\Big(\!\max\!\Big\{\phi_{k,i}\!\left(r\right)\!, \varphi_{k,i}\!\left(r\right)\!\Big\}\Big)
	f_{\!R_k}\!\left(r\right)\,\!\mathrm{d}r.
\end{align*}
Here,~(a) follows from the independence of different tiers and  the property of $\varphi_{k,i}\left(\varphi_{i,k}\left(r\right)\right) = \phi_{k,i}\left(\phi_{i,k}\left(r\right)\right) = r$, and~(b) follows from the fact that $\mathbb{E}_Y \left[\mathbb{P}\left(X>Y\right)\right] = \int_0^\infty \bar{F}_X\left(y\right)f_Y\left(y\right)\, \mathrm{d}y$ for positive random variables $X$ and $Y$.
\end{remark}

\section{Performance Analysis}\label{sec:Performance}
In this section, we will analyze the network performance in terms of the SINR coverage probability, rate coverage probability, and area sum rate.

\subsection{SINR Coverage Analysis}
The SINR coverage probability $\mathcal{C}_\nu\left(\tau\right)$, $\nu\in\left\{\text{DL,UL}\right\}$, is defined as the probability that the instantaneous received SINR is greater than a threshold $\tau$, and can be described as
\begin{align}\label{eq:Ck2C}
	\mathcal{C}_\nu\left(\tau\right) = \sum_{k\in\mathcal{K}} \mathcal{A}_{\nu,k}\mathcal{C}_{\nu,k}\left(\tau\right),
\end{align}
where $\mathcal{C}_{\nu,k}\left(\tau\right)$, which is the SINR coverage probability conditioned on the typical UE being associated with the $k$th tier, can be expressed as
\begin{align}
	\mathcal{C}_{\nu,k}\left(\tau\right) 
	= \mathbb{P}\left( \left.\mathrm{SINR}_{\nu,k} > \tau \right| K_\nu = k \right)
	= \mathbb{P}\left( \left.\frac{P_{\nu,k} G_k h \ell_k\left(\lVert \mathbf{x}_\nu^* \rVert\right)}{\sigma_k^2 + I_{\nu,k}} > \tau \right| K_\nu = k \right),
\end{align}
where $P_{\nu,k}$ is the transmit power of the serving BS in the $k$th tier for $\nu=\text{DL}$ and the transmit power of the typical UE for $\nu=\text{UL}$.

In our analysis, we assume that each BS has at least one UE in its coverage area and thus all BSs are active in both downlink and uplink. To facilitate the calculation of uplink interference, we use $\Phi_u\in\Phi_\text{U}$ to denote the set of the active UEs in the uplink. We also assume that each BS has only one active UE at a given time, and the active UE is uniformly distributed in the uplink coverage area of its serving BS. Therefore it is reasonable to conclude that the uplink active UE set $\Phi_u$ is approximately distributed in the plane with density $\sum_{k=1}^{3}\lambda_k$. The exact distribution of $\Phi_u$, however, is unknown due to the dependance induced by the Voronoi tessellation \cite{TWC2013Novlan_UL,CL2017Haenggi_UL}. Here, we assume that $\Phi_u$ approximately forms a homogeneous PPP with density $\sum_{k=1}^{3}\lambda_k$, and more precisely, the UEs associated with BSs belonging to $\Phi_k$ in the uplink form a homogeneous PPP $\Phi_{u,k}$ with density $\lambda_k$ \cite{TWC2013Novlan_UL,TWC2017Bacha,TWC2017Ding_UL}. 

Before deriving the SINR coverage probability, we first present the Laplace transforms of the interference terms $I_{\text{DL},k}$ and $I_{\text{UL},k}$ in following lemma.
\begin{lemma}\label{lemma_LT}
\textit{The Laplace transforms of interference $I_{\text{DL},k}$ and $I_{\text{UL},k}$, conditioned on the typical UE being associated with the $k$th tier in downlink and uplink, are given by 
\begin{align}\label{eq:expression_DL_laplace}
	\mathscr{L}_{I_{\text{DL},k}}\left(s;x\right) \!=\!
	\begin{cases}
		\exp\left(
			-2\pi\lambda_1 V\left(\phi_{k,1}\left(x\right),\alpha,s T_1\right)
			-2\pi\lambda_2 V\left(\phi_{k,2}\left(x\right),\alpha,s T_2\right)
			\right), 
			\quad k\in\left\{1,2\right\}\\
		\exp\left\{-2\pi\lambda_3 \!\sum_{j\in\left\{\text{M,m}\right\}} \!p_j\left[
			W_\mathrm{L}\left(\phi_{k,\mathrm L}\left(x\right),\alpha_\mathrm{L},s T_\mathrm{L} \hat{G}_j\right)\right.\right. \\
			\qquad\qquad\qquad\qquad\qquad\quad \left.\left.+ W_\mathrm{N}\left(\phi_{k,\mathrm N}\left(x\right),\alpha_\mathrm{N},s T_\mathrm{N} \hat{G}_j\right)\right]
			\right\}, \quad k\in\left\{\mathrm{L,N}\right\}
	\end{cases}
\end{align}
and
\begin{align}\label{eq:expression_UL_laplace} 
	\mathscr{L}_{I_{\text{UL},k}}\left(s;x\right) \!=\!
	\begin{cases}
		\exp\left(
			-2\pi\lambda_1\int_{0}^\infty V\left(\max\left\{\varphi_{k,1}\left(\varphi_{k,1}\left(x\right)\right),u\right\},\alpha,s T'_1 u^{\epsilon\alpha}\right) f_{\mathcal{X}_{\mathrm{UL},1}}\left(u\right)\,\mathrm{d}u\right.\\
		\ {\left.
			-2\pi\lambda_2\int_{0}^\infty V\left(\max\left\{\varphi_{k,2}\left(\varphi_{k,2}\left(x\right)\right),u\right\},\alpha,s T'_2 u^{\epsilon\alpha}\right) f_{\mathcal{X}_{\mathrm{UL},2}}\left(u\right)\,\mathrm{d}u
			\right)}, 
		\ k\in\left\{1,2\right\}\\
		\exp\left\{-2\pi\lambda_3 \!\sum_{j\in\left\{\text{M,m}\right\}} \!p_j\left[
			W_\mathrm{L}\left(\varphi_{k,\mathrm L}\left(x\right),\alpha_\mathrm{L},s T'_\mathrm{L} \hat{G}_j\right)\right.\right.\\
		\quad\qquad\qquad\qquad\qquad\qquad\qquad \left.\left. + W_\mathrm{N}\left(\varphi_{k,\mathrm N}\left(x\right),\alpha_\mathrm{N},s T'_\mathrm{N} \hat{G}_j\right)\right]
			\right\}, 
		\ k\in\left\{\mathrm{L,N}\right\}
	\end{cases}
\end{align}
respectively, where
\begin{align*}
	V\left(x,\alpha,\beta\right) \!&=\! \frac{\beta x^{-\alpha+2}}{\alpha-2}{}_2F_1\left[1,1-\frac{2}{\alpha};2-\frac{2}{\alpha};-\beta x^{-\alpha}\right],\\
	W_\mathrm{L}\left(x,\alpha,\beta\right) &= \int_{x}^{\infty} \frac{r}{1 + \beta^{-1}r^\alpha}P_\mathrm{L}\left(r\right)\,\mathrm{d}r,\\
	W_\mathrm{N}\left(x,\alpha,\beta\right) &= \int_{x}^{\infty} \frac{r}{1 + \beta^{-1}r^\alpha}P_\mathrm{N}\left(r\right)\,\mathrm{d}r,
\end{align*}
with $\hat{G}_\mathrm{M}=1$, $\hat{G}_\mathrm{m}={G_\mathrm{m}}/{G_\mathrm{M}}$, and ${}_2F_1\left[\cdot\right]$ denoting the hypergeometric function.}
\end{lemma}
\begin{IEEEproof}
See Appendix \ref{appendix:LT}.
\end{IEEEproof}

Based on Lemma \ref{lemma_LT}, we now present the SINR coverage probability in the following theorem.
\begin{theorem}\label{theorem_sinrcoverage}
\textit{The SINR coverage probability is given by}
\setcounter{equation}{20}
\begin{align}\label{eq:sinrcov}
	\mathcal{C}_\nu\left(\tau\right) = \sum_{k\in\mathcal{K}}
	\mathcal{A}_{\nu,k}
	\int_0^\infty \exp\left(-\frac{\tau\sigma_k^2}{S_{\nu,k}\left(x\right)}\right) \cdot\mathscr{L}_{I_{\nu,k}}\left(\frac{\tau}{S_{\nu,k}\left(x\right)};x\right)
	f_{\mathcal{X}_{\nu,k}}\!\left(x\right)\!\, \mathrm{d}x,
\end{align}
\textit{where 
$\nu\in\left\{\text{DL,UL}\right\}$, 
$S_{\nu,k}\left(x\right) = P_{\nu,k} G_k \ell_k\left(x\right)$, 
$f_{\mathcal{X}_{\nu,k}}\left(r\right)$ is given by Lemma \ref{corollary_condistence}, 
and $\mathscr{L}_{I_{\nu,k}}\left(s;x\right)$ is given by Lemma \ref{lemma_LT}.}
\end{theorem}
\begin{IEEEproof}
See Appendix \ref{appendix:sinrcoverage}.
\end{IEEEproof}

\begin{remark}\label{remark:theorem1}
As can be seen from Theorem~\ref{theorem_sinrcoverage}, the distribution of the minimum distance $R_k$, the distribution of the conditional serving distance $\mathcal{X}_{\nu,k}$, and the interference $I_{\nu,k}$ play active roles in determining the value of $\mathcal{C}_\nu\left(\tau\right)$, and their impacts on the network performance will be shown in Section~\ref{sec:Simulation}. Moreover, it is noticed that a double integral is required for the calculation of $\mathcal{C}_\text{DL}\left(\tau\right)$ and a triple integral is required for $\mathcal{C}_\text{UL}\left(\tau\right)$.
\end{remark}

\begin{remark}\label{remark:couple}
Theorem~\ref{theorem_sinrcoverage} gives the downlink and uplink SINR coverage probability with decoupled access. As a special case, the uplink SINR coverage probability with coupled access can be easily derived by replacing $\mathcal{A}_{\text{UL},k}$ and $\varphi_{k,i}\left(x\right)$ in \eqref{eq:sinrcov} with $\mathcal{A}_{\text{DL},k}$ and $\phi_{k,i}\left(x\right)$, respectively, and is given by 
\begin{align}\label{eq:sinrcov_couple}
	&\mathcal{C}_\text{UL}^\text{couple}\left(\tau\right) = \sum_{k\in\mathcal{K}}
	\mathcal{A}_{\text{DL},k}
	\int_0^\infty \exp\left(-\frac{\tau\sigma_k^2}{S_{\text{UL},k}\left(x\right)}\right) \cdot\mathscr{L}_{I'_{\text{UL},k}}\left(\frac{\tau}{S_{\text{UL},k}\left(x\right)};x\right)
	f_{\mathcal{X}_{\text{DL},k}}\!\left(x\right)\!\, \mathrm{d}x,
\end{align}
where 
\begin{align}
\mathscr{L}_{I'_{\text{UL},k}}\left(s;x\right) = \left\{ \mathscr{L}_{I_{\text{UL},k}}\left(s;x\right) \left| { \varphi_{k,i}\left(r\right) = \left(\frac{B_i T_i}{B_k T_k}\right)^\frac{1}{\alpha_i} r^\frac{\alpha_k}{\alpha_i} } \right.\right\}.\nonumber
\end{align}
\end{remark}

\subsubsection{Sparse BS Case}
When the BSs are sparse, i.e., $\sum_{k=1}^3\lambda_k \ll \lambda_\text{U}$, the interference can be neglectable, which means that the network is noise limited. Therefore, the SINR coverage probability in Theorem~\ref{theorem_sinrcoverage} can be reduced to signal-to-noise ratio~(SNR) coverage probability by setting the interference $I_{\nu,k}$ to zero, namely
\begin{align}
	&\mathcal{C}_\nu\left(\tau\right) \!=\! \sum_{k\in\mathcal{K}}
	\mathcal{A}_{\nu,k}\!
	\int_0^\infty\!
	\exp\left(-\frac{\tau\sigma_k^2}{S_{\nu,k}\left(x\right)}\right)\! f_{\mathcal{X}_{\nu,k}}\!\left(x\right)\, \mathrm{d}x.
\end{align}

\subsubsection{Dense BS Case}
Due to the increased demands of mobile data traffic, network densification is considered as a key mechanism in the evolution of cellular networks \cite{CM2017ultradense}. Network densification shortens the distance from UEs to BSs, and the network becomes interference limited. With the increase of network density, some BSs may not serve any UEs. 
In such a scenario, the density of the downlink active BSs of the $k$th tier is $\lambda_k^\prime = \min\left\{\lambda_k,\lambda_\text{U}\mathcal{A}_{\text{DL},k}\right\}$, and the density of the uplink active UEs associated with the $k$th tier BSs is $\lambda_{u,k}^\prime = \min\left\{\lambda_k,\lambda_\text{U}\mathcal{A}_{\text{UL},k}\right\}$. By replacing $\lambda_k$ in $\mathscr{L}_{I_{\text{DL},k}}\left(s;x\right)$ and $\mathscr{L}_{I_{\text{UL},k}}\left(s;x\right)$ with $\lambda_k^\prime$ and $\lambda_{u,k}^\prime$, respectively, and setting $\sigma_k^2 = 0$, we can obtain the signal-to-interference ratio (SIR) coverage probability for dense BS case, which is given by
\begin{align}
	&\mathcal{C}_\nu\left(\tau\right) = \sum_{k\in\mathcal{K}}
	\mathcal{A}_{\nu,k}
	\int_0^\infty 
	\mathscr{L}_{I_{\nu,k}^\prime}\left(\frac{\tau}{S_{\nu,k}\left(x\right)};x\right) f_{\mathcal{X}_{\nu,k}}\left(x\right)\, \mathrm{d}x,
\end{align}
where
$\mathscr{L}_{I_{\text{DL},k}^\prime}\left(s;x\right) = \left\{ \left.\mathscr{L}_{I_{\text{DL},k}}\left(s;x\right) \right|{\lambda_k = \min\left\{\lambda_k,\lambda_\text{U}\mathcal{A}_{\text{DL},k}\right\} } \right\}$ and
$\mathscr{L}_{I_{\text{UL},k}^\prime}\left(s;x\right) = \left\{ \left.\mathscr{L}_{I_{\text{UL},k}}\left(s;x\right) \right|\right.$\\
$\left.\lambda_k = \min\left\{\lambda_k,\lambda_\text{U}\mathcal{A}_{\text{UL},k}\right\} \right\}$.

\subsection{Rate Coverage Probability}
To quantify the uplink performance improvement with decoupled access, the user-perceived rate coverage probability $\mathcal{R}\left(\rho\right)$, which is defined as the probability of the instantaneous UE data rate being higher than a threshold $\rho$, is presented here as a relevant metric. Compared with SINR coverage probability, the effect of cell load is taken into account in the rate coverage probability. 
According to \cite{TWC2013Singh_load,TWC2016De_Elshaer}, the approximate mean load $N_{\nu,k}$ of each BS in the $k$th tier is given by
\begin{align}\label{eq:load}
	N_{\nu,k} = 1 + \frac{1.28 \mathcal{A}_{\nu,k} \lambda_\text{U}}{\lambda_k}.
\end{align}
Therefore, the user-perceived rate coverage probability $\mathcal{R}_\nu\left(\rho\right)$ can be formulated as
\begin{align}\label{eq:ratecov_prior}
	\mathcal{R}_\nu\left(\rho\right)
	&= \sum_{k\in\mathcal{K}} \mathcal{A}_{\nu,k} \mathbb{P}\left[ \frac{W_k \log_2\left(1+\text{SINR}_{\nu,k}\right)}{N_{\nu,k}} > \rho \right],
\end{align}
where $W_k$ is the carrier bandwidth. Leveraging the SINR coverage probability in Theorem~\ref{theorem_sinrcoverage}, the expression of $\mathcal{R}_\nu\left(\rho\right)$ is given in the following theorem.

\begin{theorem}\label{theorem_ratecoverage}
\textit{The user-perceived rate coverage probability $\mathcal{R}_\nu\left(\rho\right)$ is given by}
\begin{align}
	&\mathcal{R}_\nu\left(\rho\right) = \sum_{k\in\mathcal{K}} \mathcal{A}_{\nu,k} \mathcal{C}_{\nu,k}\left(
	2^\frac{\rho N_{\nu,k}}{W_k} - 1
	\right).
\end{align}
\end{theorem}
\begin{IEEEproof}
The proof can be easily derived from \eqref{eq:ratecov_prior}.
\end{IEEEproof}

\subsection{Area Sum Rate}
The ASR, denoting the sum throughput normalized by the area with unit of $\text{bps/km}^2$, can be described as
\begin{align}\label{eq:ASR_define}
	\text{ASR}_\nu 
	&= \frac{\mathbb{E}\left[\sum_{k\in\mathcal{K}} n_k W_k \log_2\left(1+\text{SINR}_{\nu,k}\right)\right]}{\lvert S \rvert},
\end{align}
where $\lvert S \rvert$ is the area of $S$, $n_k$ is the number of BSs located in $S$. And the expression of $\text{ASR}_\nu$ is givn in the following theorem.

\begin{theorem}\label{theorem_ASR}
\textit{The network ASR is given by
\begin{align}
	&\text{ASR}_\nu = \sum_{k\in\mathcal{K}} \gamma_k W_k
	\int_0^\infty \int_0^\infty \exp\left(-\frac{\left(2^\rho-1\right)\sigma_k^2}{S_{\nu,k}\left(x\right)}\right) \cdot\mathscr{L}_{I_{\nu,k}}\left(\frac{2^\rho-1}{S_{\nu,k}\left(x\right)};x\right) f_{\mathcal{X}_{\nu,k}}\left(x\right)\, \mathrm{d}x\, \mathrm{d}\rho.
\end{align}
where $\gamma_k=\lambda_k$ for $k\in\left\{1,2\right\}$ and $\gamma_k=\frac{\lambda_k\mathcal{A}_{\nu,k}}{\mathcal{A}_{\nu,\mathrm{L}}+\mathcal{A}_{\nu,\mathrm{N}}}$ for $k\in\left\{\mathrm{L,N}\right\}$. }
\end{theorem}
\begin{IEEEproof}
From \eqref{eq:ASR_define}, we have
\begin{align}\label{proof_ASR}
	\text{ASR}_\nu &= \sum_{k\in\left\{1,2\right\}} \lambda_k W_k \mathbb{E} \left[\log_2\left(1+\mathrm{SINR}_{\nu,k}\right)\right] 
	+ \sum_{k\in\left\{\mathrm{L,N}\right\}} \frac{\lambda_k\mathcal{A}_{\nu,k}}{\mathcal{A}_{\nu,\mathrm{L}}+\mathcal{A}_{\nu,\mathrm{N}}}  W_k \mathbb{E} \left[\log_2\left(1+\mathrm{SINR}_{\nu,k}\right)\right] \nonumber\\
	&\overset{(a)}{=} \sum_{k\in\mathcal{K}} \gamma_k W_k \int_0^\infty \mathbb{P}\left[ \log_2\left(1+\mathrm{SINR}_{\nu,k}\right) > \rho \right]\, \mathrm{d}\rho\nonumber\\
	&= \sum_{k\in\mathcal{K}} \gamma_k W_k \int_0^\infty \mathcal{C}_{\nu,k}\left(2^\rho - 1\right)\, \mathrm{d}\rho,
\end{align}
where (a) follows from $\mathbb{E}\left[X\right] = \int_0^\infty \mathbb{P}\left[X>x\right]\, \mathrm{d}x$ for positive random variable $X$. Plugging $\mathcal{C}_{\nu,k}\left(\tau\right)$ from Theorem~\ref{theorem_sinrcoverage} into \eqref{proof_ASR}, we can obtain the expression of $\text{ASR}_\nu$.
\end{IEEEproof}

Since ASR takes the cells density into account, it can be used to describe the network performance gain induced by the network densification.

\begin{table}[!t]
\renewcommand{\arraystretch}{1.3}
\caption{Notations and Default Simulation Values}
\label{table:parameter}
\centering
\begin{tabular}{p{5em}p{27em}p{9em}}
\hline
\bfseries Notation & \bfseries Description & \bfseries Value\\
\hline
$\lambda_1,\lambda_2,\lambda_3$						& Density of sub-6GHz MCells PPP $\Phi_1$, sub-6GHz SCells PPP $\Phi_2$, and mmWave SCells PPP $\Phi_3$, respectively. 	& $5\,\text{/km}^2$,  $30\,\text{/km}^2$,  $30\,\text{/km}^2$\\
\hline
$P_\text{DL,1}$,$P_\text{DL,2}$,\ $P_\text{DL,3}$	& Downlink transmit power of sub-6GHz MCells, sub-6GHz SCells, and mmWave SCells, respectively. 	& $46\,\text{dBm}$,  $40\,\text{dBm}$,  $30\,\text{dBm}$\\
\hline
$\lambda_\mathrm{U}$								& Density of UEs PPP $\Phi_\text{U}$. 	& $200\,\text{km}^2$\\
\hline
$P_u$												& Uplink initial transmit power of UEs before applying power control. & $23\,\text{dBm}$\\
\hline
$\epsilon$											& Uplink power control fraction for sub-6GHz transmission. & $0.2$ \\
\hline
$G_\mathrm{M},G_\mathrm{m},\theta_\mathrm{b}$		& Main-lobe gain, side-lobe gain and main-lobe beamwidth of the sectored antenna model, respectively. & $18\,\text{dBi}$, $-2\,\text{dBi}$, $10^\circ$\\
\hline
$f_\mathrm{S},f_\mathrm{M}$							& Sub-6GHz and mmWave system carrier frequencies, respectively. & $2\,\text{GHz}$, $28\,\text{GHz}$\\
\hline
$W_s,W_m$				& Sub-6GHz system bandwidth and mmWave system bandwidth, respectively. & $20\,\text{MHz}$, $1\,\text{GHz}$\\
\hline
$C_\mathrm{S},C_\mathrm{L},C_\mathrm{N}$			& Path loss intercepts for sub-6GHz, mmWave LoS and mmWave NLoS signals, respectively. & $-38.5 \text{dB}$, $-61.4 \text{dB}$, $-72 \text{dB}$ \cite{JSAC2014Akdeniz} \\
\hline
$\alpha,\alpha_\mathrm{L},\alpha_\mathrm{N}$		& Path loss exponents for sub-6GHz, mmWave LoS and mmWave NLoS signals, respectively. & $3$,  $2$,  $2.92$ \cite{JSAC2014Akdeniz}\\
\hline
$p_\mathrm{L},R_\mathrm{B}$							& Parameters in generalized blockage ball model. & $0.2$, $200\,\text{m}$\\
\hline
$\sigma^2$ 											& $-174\,\text{dBm/Hz} + 10\log_{10}\left(W\right) + 10\,\text{dB}$ 		& \\
\hline
\end{tabular}
\end{table}

\section{Results and Discussions}\label{sec:Simulation}
In this section, Monte Carlo simulations are conducted to validate the accuracy of our theoretical analysis and to investigate the effects of different factors on the network performance. For convenience, Table \ref{table:parameter} summarizes the notations used in this paper together with the default values employed in the simulations.

\subsection{Association Probability}
\begin{figure}[!t]
	\centering
	\subfloat[]
	{\includegraphics[scale=0.5]{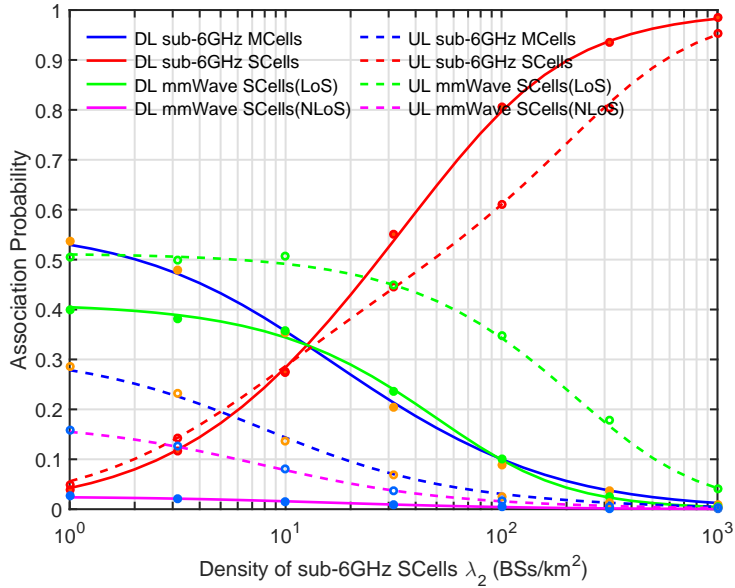}\label{fig1a_association_lambda2}}
	\hfil
	\subfloat[]
	{\includegraphics[scale=0.5]{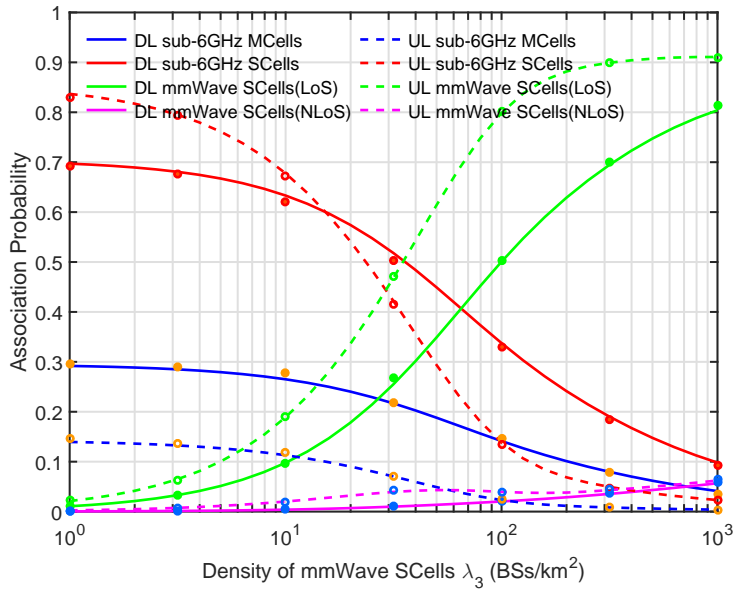}\label{fig1b_association_lambda3}}
	\caption{Downlink and uplink association probability vs. the densities of sub-6GHz/mmWave SCells. The density varies from $1\text{/km}^2$ to $1000\text{/km}^2$. The circles represent the corresponding simulation results.}
	\label{association_lambda}
\end{figure}
\begin{figure}[!t]
	\centering
	\includegraphics[scale=0.5]{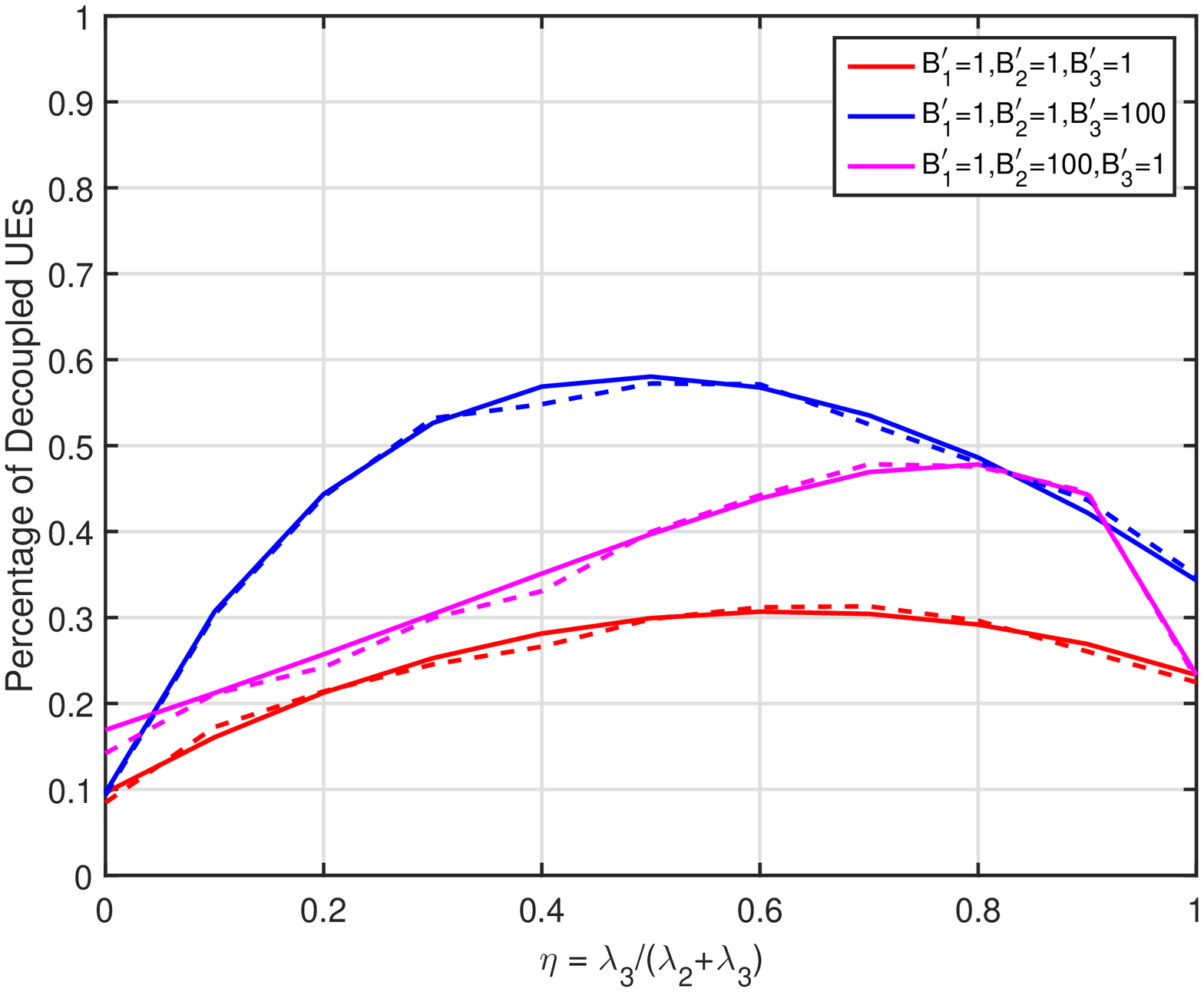}
	\caption{Percentage of decoupled UEs $\mathcal{D}$ vs. the fraction of mmWave SCells $\eta$ with $\lambda_1=5\text{/km}^2$ and $\lambda_2+\lambda_3=60\text{/km}^2$. The dashed lines represent the simulation results.}
	\label{association_decouplePercent}
\end{figure}
The association probability with variable SCell density is shown in Fig.~\ref{association_lambda}. It can be observed that the analytical and simulation results match very well. The difference between the downlink and uplink association probabilities shows the impact of decoupled access. In mmWave SCell scenario, the uplink association probability is higher than the downlink one, and this is due to the fact that the uplink coverage of mmWave SCells is usually larger than the downlink coverage, which is in line with the results of \cite{TWC2016De_Elshaer}. With the densifying of sub-6GHz and mmWave SCells, the association probabilities of sub-6GHz and mmWave SCells monotonically increase, as shown in Figs.~\ref{fig1a_association_lambda2} and \ref{fig1b_association_lambda3}, respectively, which can be explained by the fact that the higher the density of SCells, the lower the minimum distance from UEs to SCells, and hence the larger the received signal power. As a result, the traffic can be offloaded from MCells to SCells efficiently.

The percentage of decoupled UEs $\mathcal{D}$ with the fraction of mmWave SCells $\eta$ is shown in Fig.~\ref{association_decouplePercent}. It can be seen that the analytical results match well with the simulation ones. The percentage of decoupled UEs initially increase and then decrease with the growing of $\eta$, which indicates that the more heterogeneous the network is, the more evident the DUDe will be. Moreover, it can be seen that the uplink bias factor $B^\prime_k$ has a significant effect on $\mathcal{D}$. The higher $B^\prime_k \left(k\in\left\{2,3\right\}\right)$ is, the more UEs will be associated with SCell, and thus the higher $\mathcal{D}$ will be.

\subsection{Coverage Results}
\subsubsection{SINR Coverage Probability}
\begin{figure}[!t]
	\centering
	\subfloat[Sparse case]
	{\includegraphics[scale=0.5]{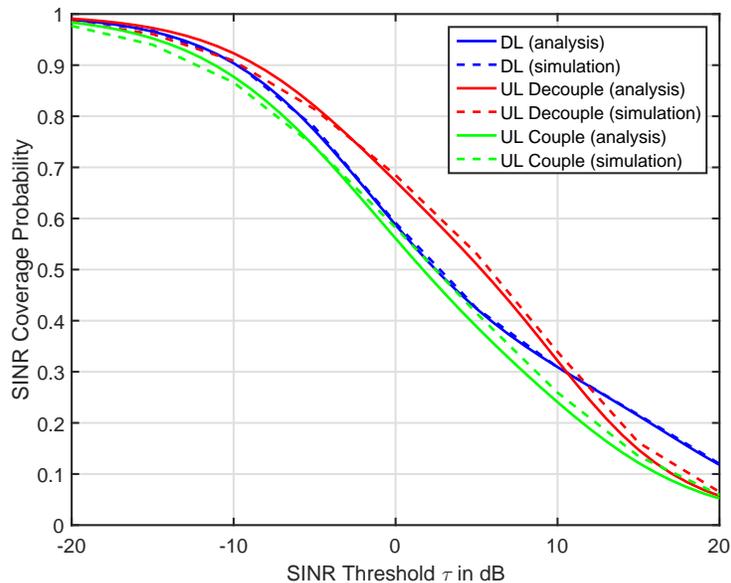}\label{fig2a_sinrcov_thr}}
	\hfil
	\subfloat[Dense case]
	{\includegraphics[scale=0.5]{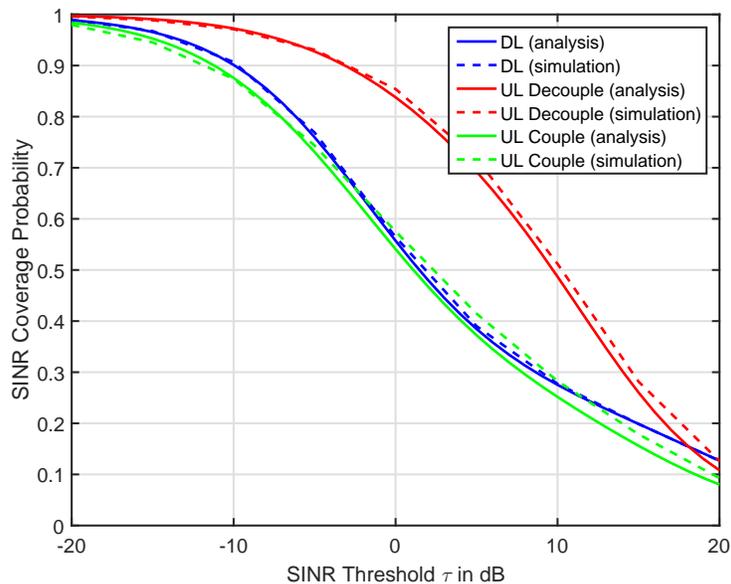}\label{fig2b_sinrcov_thr}}
	\caption{Downlink and uplink SINR coverage probability vs. the SINR threshold $\tau$ in sparse/dense cases. The parameters are selected as $\lambda_1=\text{5/km}^2$, $\lambda_2=\text{30/km}^2$, $\lambda_3=\text{30/km}^2$, $\lambda_\text{UE}=\text{200/km}^2$ in sparse case~(a), and $\lambda_1=\text{15/km}^2$, $\lambda_2=\text{100/km}^2$, $\lambda_3=\text{100/km}^2$, $\lambda_\text{UE}=\text{500/km}^2$ in dense case~(b).}
	\label{sinrcov_thr}
\end{figure}
The SINR coverage probabilities for sparse and dense network cases are presented in Figs.~\ref{fig2a_sinrcov_thr} and \ref{fig2b_sinrcov_thr}, respectively. We observe that the simulation curves are a bit higher than the analytical ones, which can be explained by the following two reasons. First, the independence assumption of BSs and active UEs used in theoretical analysis ignores the correlation of the transmitters' locations, and thus the analytical results represent the worst coverage scenario. Second, it is unrealistic to ensure all BSs are active in Monte Carlo simulations, i.e., there are a few BSs with no UEs to serve, the interference in simulations is generally less than that in theoretical analysis, which will lead to higher coverage probability in simulation results. Moreover, the uplink SINR coverage probability with decoupled access is higher than that with coupled access, especially for the dense case, which indicates that DUDe could improve the network's uplink performance efficiently \cite{CM2016DUDe}.


\subsubsection{Rate Coverage Probability}
\begin{figure}[!t]
	\centering
	\subfloat[Sparse case]
	{\includegraphics[scale=0.5]{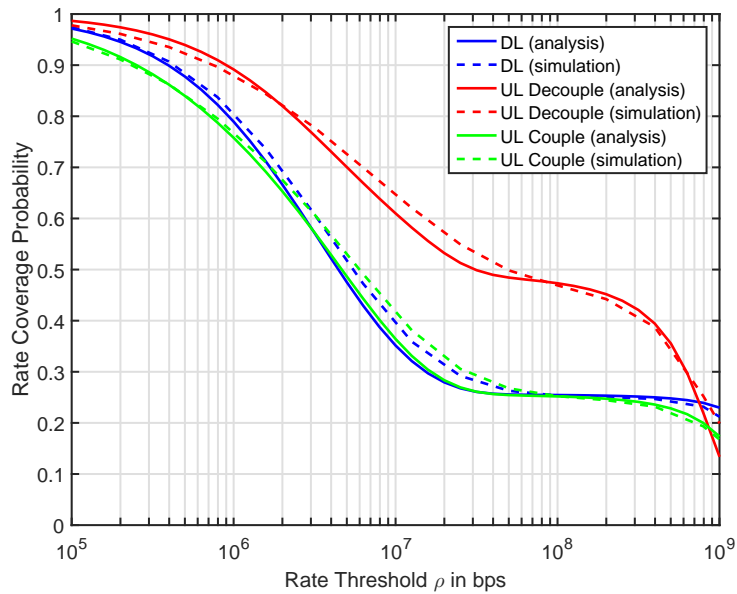}\label{fig3a_ratecov_thr}}
	\hfil
	\subfloat[Dense case]
	{\includegraphics[scale=0.5]{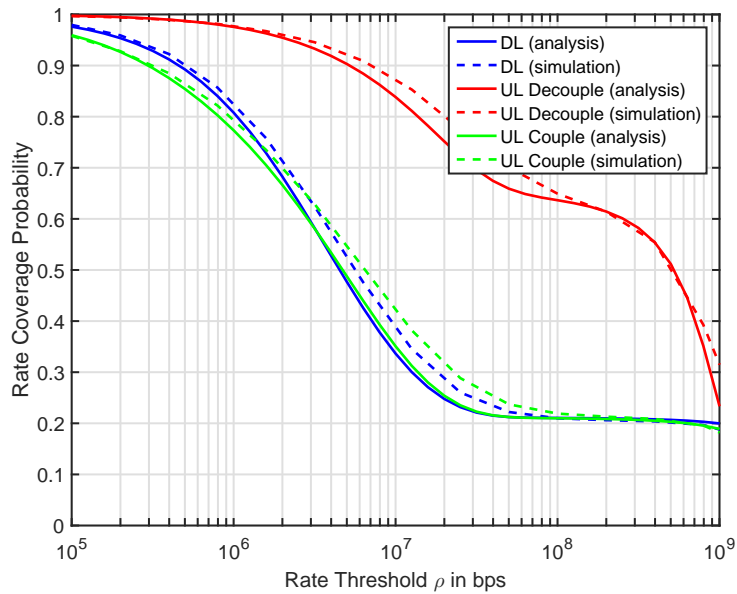}\label{fig3b_ratecov_thr}}
	\caption{Downlink and uplink user-perceived rate coverage probability vs. the rate threshold $\rho$ in sparse/dense cases. The parameters are selected as $\lambda_1=\text{5/km}^2$, $\lambda_2=\text{30/km}^2$, $\lambda_3=\text{30/km}^2$, $\lambda_\text{UE}=\text{200/km}^2$ in sparse case~(a), and $\lambda_1=\text{15/km}^2$, $\lambda_2=\text{100/km}^2$, $\lambda_3=\text{100/km}^2$, $\lambda_\text{UE}=\text{500/km}^2$ in dense case~(b).}
	\label{ratecov_thr}
\end{figure}
\begin{figure}[!t]
	\centering
	\subfloat[]
	{\includegraphics[scale=0.5]{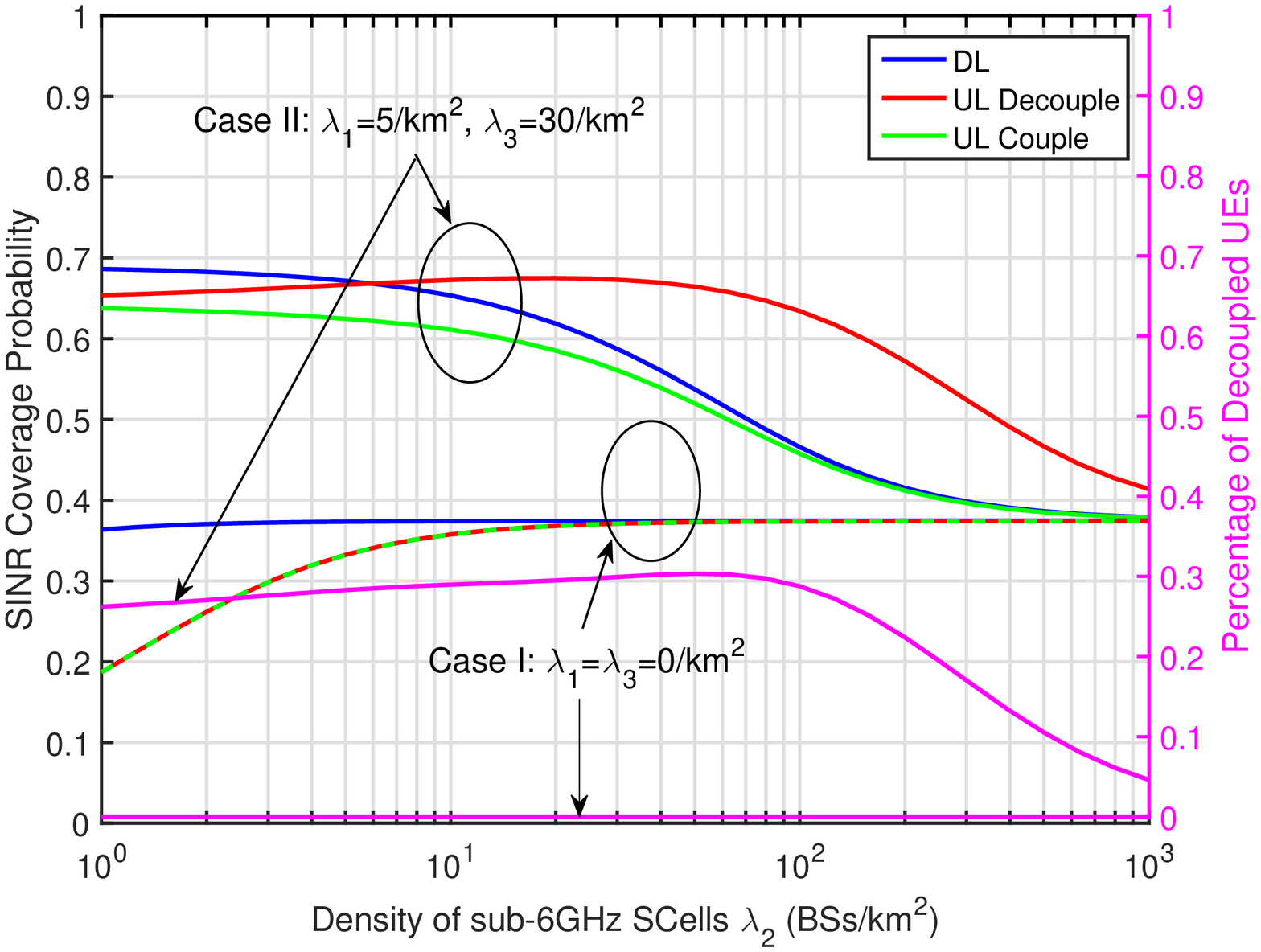}\label{fig4a_sinrcov_lambda2}}
	\hfil
	\subfloat[]
	{\includegraphics[scale=0.5]{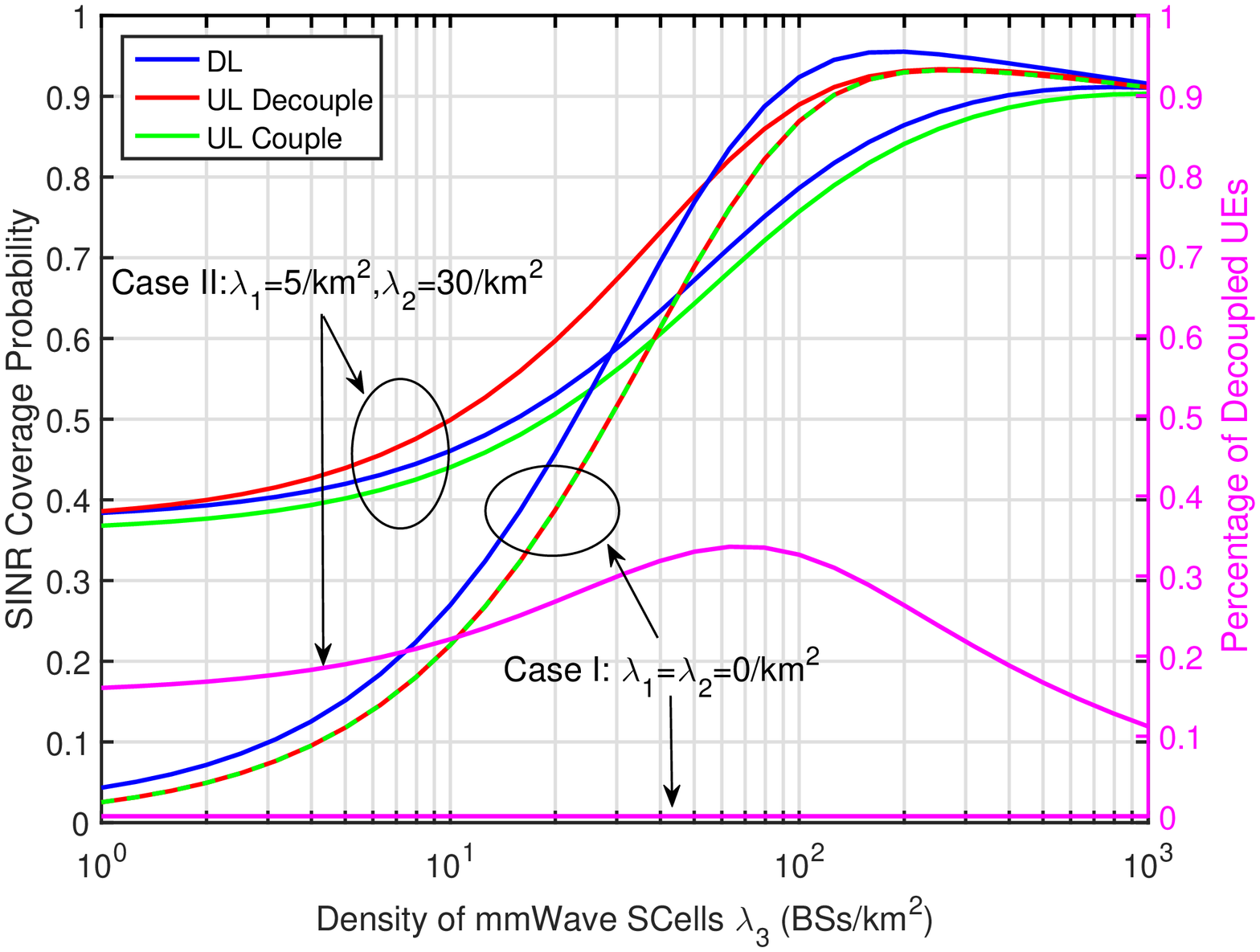}\label{fig4b_sinrcov_lambda3}}
	\caption{Downlink and uplink SINR coverage probability vs. the density of SCells with $\tau=0$ dB.}
	\label{sinrcov_lambda}
\end{figure}
The curves of user-perceived rate coverage probabilities for sparse and dense network cases are plotted in Figs.~\ref{fig3a_ratecov_thr} and \ref{fig3b_ratecov_thr}, respectively. 
The uplink rate coverage probability $\mathcal{R}_\text{UL}\left(\rho\right)$ with decoupled access is evidently much higher than $\mathcal{R}_\text{UL}\left(\rho\right)$ with coupled access, especially for the dense case in Fig.~\ref{fig3b_ratecov_thr}. These observations indicate that DUDe could improve the network's uplink performance significantly. 

The decoupled gain of uplink performance mainly benefits from the large available bandwidth of mmWave. As can be seen from Fig.~\ref{fig1b_association_lambda3}, the UEs are more likely to be associated with mmWave SCells after applying decoupled access. Since the bandwidth of sub-6GHz and mmWave are selected as $W_\text{sub-6GHz}=20\,\text{MHz}$ and $W_\text{mmWave}=1\,\text{GHz}$ here, the uplink average rate that proportional to the bandwidth will be improved significantly. Moreover, there is a flat area around $\rho=10^8$ bps. This is because that $\mathcal{R}\left(\rho\right) = \sum_{k\in\mathcal{K}}\mathcal{A}_k \mathcal{R}_k\left(\rho\right)$ is a weighted sum of $\mathcal{R}_k\left(\rho\right)$, and the $\mathcal{R}_k\left(\rho\right)$ of mmWave band is much higher than that of sub-6GHz.

\subsection{Network Performance Trends}
Here, we will investigate the impact of SCells densification on the network performance including SINR coverage probability and ASR. The decoupled access is also addressed in contrast with coupled access.

\subsubsection{SINR Coverage Trends}
The SINR coverage probability curves with different sub-6GHz and mmWave SCell densities are presented in Figs.~\ref{fig4a_sinrcov_lambda2} and \ref{fig4b_sinrcov_lambda3}, respectively. 
In case I of Fig.~\ref{fig4a_sinrcov_lambda2}, the downlink SINR coverage probability almost remains unchanged, which can be explained by the fact the both the desired signal and interference increase with the densification of sub-6GHz SCells densification, and the uplink SINR coverage probability initially grow as $\lambda_2$ increases and eventually remains unchanged, which is due to the different increase speeds of desired signal and interference powers with the increase of $\lambda_2$. In case II of Fig.~\ref{fig4a_sinrcov_lambda2}, both the downlink and uplink SINR coverage probabilities with coupled access monotonically decrease as the interference increases with the the increase of $\lambda_2$, and the uplink SINR coverage probability with decoupled access first slightly increases and then decreases with the increase of $\lambda_2$.

Comparing the two cases in Fig.~\ref{fig4a_sinrcov_lambda2}, we can find that the adding of sub-6GHz MCells and mmWave SCells boosts the SINR coverage probability, especially when the network is sparse. Moreover, the uplink SINR coverage probability with decoupled access is higher than that with coupled access in case II, but there is no evident differences between them in case I, which implies that DUDe is preferred in HetNets. 
It is noticed that the SINR coverage probabilities of case II converge to that of case I with the increase of $\lambda_2$, which is due to the fact that sub-6GHz SCells dominate the network performance with extremely high $\lambda_2$.

In Fig.~\ref{fig4b_sinrcov_lambda3}, it can be seen that both downlink and uplink SINR coverage probabilities increase with $\lambda_3\in\left[1,100\right]/\text{km}^2$ and then suffer from a slow growth~(for case II) or even a decrease~(for case I) with $\lambda_3\in\left[100,1000\right]/\text{km}^2$. This can be explained as followed. Since mmWave SCells are noise limited in sparse case, adding more mmWave SCells will increase the network performance. But when the network is dense enough, the receivers will suffer from high interference that counteract the improvement of desired signals. Comparing the two cases, we  find that the adding of sub-6GHz MCells and SCells boosts the uplink SINR coverage probability of coupled access with $\lambda_3\in\left[1,40\right]/\text{km}^2$ and that of decoupled access with $\lambda_3\in\left[1,100\right]/\text{km}^2$. 

Furthermore, Figs.~\ref{fig4a_sinrcov_lambda2} and \ref{fig4b_sinrcov_lambda3} show several differences between sub-6GHz and mmWave SCells. Comparing case I of Fig.~\ref{fig4a_sinrcov_lambda2} with case I of Fig.~\ref{fig4b_sinrcov_lambda3}, the sub-6GHz SCells could achieve higher SINR coverage probability when $\lambda < \text{20/km}^2$. However, the saturation occurs earlier for sub-6GHz SCells, and the mmWave network will achieve higher SINR coverage probability in dense case. 

\begin{figure}[!t]
	\centering
	\subfloat[]
	{\includegraphics[scale=0.5]{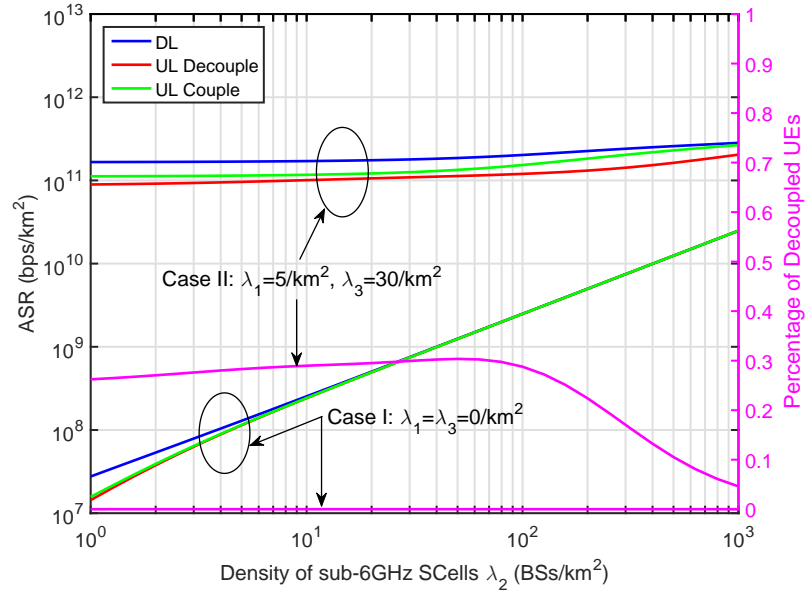}\label{fig5a_ASR_lambda2}}
	\hfil
	\subfloat[]
	{\includegraphics[scale=0.5]{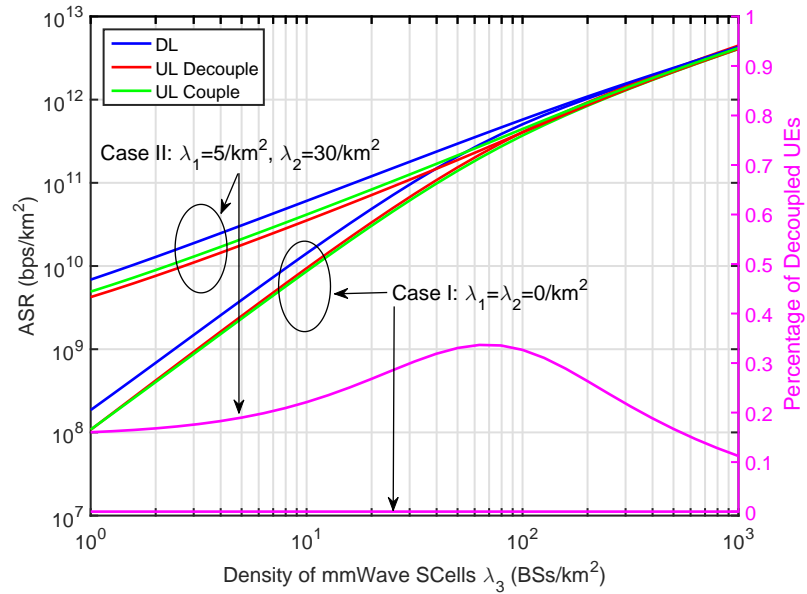}\label{fig5b_ASR_lambda3}}
	\caption{Downlink and uplink ASR vs. the density of SCells.}
	\label{ASR_lambda}
\end{figure}
\subsubsection{ASR Trends}
The ASR performance against the densities of different SCells is presented in Fig.~\ref{ASR_lambda}. 
In case I of Fig.~\ref{fig5a_ASR_lambda2}, the ASR linearly increases with $\lambda_2$. This is due to the fact that densifying sub-6GHz SCells will provide more available BSs and thus boosts the average throughput per unit area. However, in case II of Fig.~\ref{fig5a_ASR_lambda2}, the ASR remains almost unchanged. This is because the parameters in case II are set to be $\lambda_3=\text{30/km}^2$, $W_\text{sub-6GHz}=20\,\text{MHz}$, $W_\text{mmWave}=1\,\text{GHz}$, and the sub-6GHz SCells are not dense enough to provide comparable throughput comparing with the mmWave SCells. 
In Fig.~\ref{fig5b_ASR_lambda3}, the ASRs of the both cases increase with $\lambda_3$ due to the large available bandwidth of mmWave. 
The ASR is mainly dominated by the densities of cells, and it is better to deploy more SCells to achieve higher ASR.

Furthermore, it can be observed from Fig.~\ref{ASR_lambda} that the uplink ASR has no difference between coupled and decoupled access in case I, since UEs are associated with the same BSs in decoupled access when there is only one tier BSs. However, in case II, the uplink ASR with decoupled access is even a bit smaller than that with coupled access. 
This is because the decoupled access expands the uplink coverage region of mmWave SCells, which leads to the changes in cell load and SINR performance of different cells. With decoupled access, mmWave SCells are preferred in uplink, which results in that the received SINR of mmWave SCells will suffer from the increased mmWave interference and leads to the decrease of uplink ASR.

\section{Conclusion}\label{sec:Conclusion}
In this paper, we have proposed a general analytical framework to analyze the system performance of the decoupled HetNets consisting of sub-6GHz MCells, sub-6GHz SCells and mmWave SCells. The metric of ASR has been proposed to investigate the system performance of the hybrid frequency networks. We have derived the expressions for the percentage of decoupled UEs, the SINR coverage probability, the user-perceived rate coverage probability, and ASR, and have investigated the effect of SCells densification on the network performance. Applying decoupled association could significantly boost the uplink SINR and user-perceived rate coverage probabilities. To achieve high capacity and coverage probability, the density of different kind of cells need to be deployed properly. 
The accuracy of our analysis has been validated through Monte Carlo simulations.


\appendices

\section{Proof of Theorem~\ref{theorem_association}}\label{appendix:association}
The typical UE is associated with the $k$th tier in downlink if and only if $B_k T_k R_k^{-\alpha_k} > B_i T_i R_i^{-\alpha_i}$, $\forall i\in\mathcal{K}\backslash k$. Thus the downlink association probability of tier $k$ can be formulated as
\begin{align}\label{eq:app_ADLk}
	\mathcal{A}_{\text{DL},k}
	&= \mathbb{P}\left(B_k T_k R_k^{-\alpha_k} > \bigcup_{ i\in\mathcal{K}\backslash k } B_i T_i R_i^{-\alpha_i}\right) \nonumber\\
	&= \int_0^\infty \prod_{ i\in\mathcal{K}\backslash k } 
		\mathbb{P}\left(R_i > \left(\frac{B_i T_i}{B_k T_k}\right)^\frac{1}{\alpha_i} \!r^\frac{\alpha_k}{\alpha_i}\right) 
		\!f_{R_k}\left(r\right)\, \mathrm{d}r \nonumber\\
	&= \int_0^\infty \prod_{ i\in\mathcal{K}\backslash k } 
		\bar{F}_{R_i}\left[ \phi_{k,i}\left(r\right) \right] 
		\!f_{R_k}\left(r\right)\, \mathrm{d}r,
\end{align}
where $\bar{F}_{R_k}\left(r\right)$ is the complementary cumulative distribution function of $R_k$, $f_{R_k}\left(r\right)$ is given in Lemma~\ref{lemma_mindistance}, and
$\phi_{k,i}\left(r\right) 
\!=\! \left(\frac{B_i T_i}{B_k T_k}\right)^\frac{1}{\alpha_i} 
r^\frac{\alpha_k}{\alpha_i}$
is termed as the downlink distance transfer function (DDTF). If the typical UE is associated to a BS of tier $k$ with distance $r$ in downlink, then the BSs in the $i$th tier will be farther than DDTF $\phi_{k,i}\left(r\right)$.

Similarly, the uplink association probability of tier $k$ can be formulated by
\begin{align}\label{eq:app_AULk}
	\mathcal{A}_{\text{UL},k} 
	&\!=\! \mathbb{P}\left( 
		B'_k T'_k R_k^{\left(\epsilon_k-1\right)\alpha_k} 
		\!>\! \bigcup_{ i\in\mathcal{K}\backslash k } B'_i T'_i R_i^{\left(\epsilon_i-1\right)\alpha_i} \right) \nonumber\\
	&\!=\! \int_0^\infty \prod_{ i\in\mathcal{K}\backslash k } 
		\mathbb{P}\left(
		R_i \!>\! \left(\frac{B'_i T'_i}{B'_k T'_k}\right)^\frac{1}{\left(1-\epsilon_i\right)\alpha_i} 
		\!r^\frac{\left(1-\epsilon_k\right)\alpha_k}{\left(1-\epsilon_i\right)\alpha_i} \right) 
		\!f_{R_k}\!\left(r\right)\, \mathrm{d}r \nonumber\\
	&\!=\! \int_0^\infty \prod_{ i\in\mathcal{K}\backslash k } 
		\bar{F}_{R_i}\left[ \varphi_{k,i}\left(r\right) \right] 
		f_{R_k}\!\left(r\right)\, \mathrm{d}r,
\end{align}
where 
$\varphi_{k,i}\left(r\right) 
\!=\! \left(\frac{B'_i T'_i}{B'_k T'_k}\right)^\frac{1}{\left(1-\epsilon_i\right)\alpha_i} 
r^\frac{\left(1-\epsilon_k\right)\alpha_k}{\left(1-\epsilon_i\right)\alpha_i}$ 
is termed as the uplink distance transfer function (UDTF). If the typical UE is associated to a BS of tier $k$ with distance $r$ in uplink, then the BSs in the $i$th tier will be farther than UDTF $\varphi_{k,i}\left(r\right)$. From \eqref{eq:app_ADLk} and \eqref{eq:app_AULk}, we can derive \eqref{eq:Avk}.

\section{Proof of Corollary~\ref{corollary_condistence}}\label{appendix:condistence}
Conditioned on the typical UE being associated with the BS of tier $k$, the event of $\mathcal{X}_{\nu,k}\leq x$ can be rewritten as $\left.R_k\leq x \right|_{K_\nu=k}$. Leveraging the conditional probability formula, we have
\begin{align}
	F_{\mathcal{X}_{\nu,k}}\left(x\right) 
	&= \mathbb{P}\left( \left.R_k\leq x \right| K_\nu=k \right)
	= \frac{\mathbb{P}\left( R_k \leq x, K_\nu=k \right)}{\mathbb{P}\left( K_\nu=k \right)} \nonumber\\
	&= \frac{1}{\mathcal{A}_{\nu,k}}
		\mathbb{P}\left( R_k\leq x, \bigcup_{ i\in\mathcal{K}\backslash k } R_i > \Psi_{\nu,k,i}\left(R_k\right) \right) \nonumber\\
	&= \frac{1}{\mathcal{A}_{\nu,k}}
		\int_0^x \prod_{ i\in\mathcal{K}\backslash k } \bar{F}_{R_i}\left[ \Psi_{\nu,k,i}\left(r\right) \right] 
		f_{R_k}\left(r\right)\, \mathrm{d}r,
\end{align}
where $\Psi_{\text{DL},k,i}\left(r\right)=\phi_{k,i}\left(r\right)$ and $\Psi_{\text{UL},k,i}\left(r\right)=\varphi_{k,i}\left(r\right)$. The PDF of $\mathcal{X}_{\nu,k}$ follows by taking the derivative of $F_{\mathcal{X}_{\nu,k}}\left(x\right)$ with respect to $x$, which gives \eqref{eq:corollary_PDF_fx}.

\section{Proof of Lemma~\ref{lemma_LT}}\label{appendix:LT}
We first calculate the Laplace transform of interference $I_{\text{DL},k}$. When the typical UE is associated to the sub-6GHz cell in downlink, i.e., $k\in\left\{1,2\right\}$, the interference $I_{\text{DL},k}$ comes from the BSs of tiers $1$ and $2$, and it can be written as
\begin{align}
	I_{\text{DL},k} 
	&= \sum_{i\in\left\{1,2\right\}} \sum_{\mathbf{x}\in \Phi_i\backslash \mathbf{x}_\text{DL}^*} 
		T_i h_{\mathbf{x}\!\rightarrow\mathbf{y}_0} \lVert \mathbf{x}\rVert^{-\alpha} 
	\\
	\mathscr{L}_{I_{\text{DL},k}}\left(s;x\right) 
	&= \mathbb{E}_{\Phi,h} \left[ 
		\exp\left( -s I_{\text{DL},k} \right) \right] \nonumber\\
	&= \prod_{i\in\left\{1,2\right\}} \mathbb{E}_{\Phi_i} \left\{ 
		\prod_{\mathbf{x}\in \Phi_i\backslash \mathbf{x}_\text{DL}^*}
		\mathbb{E}_{h} \left[ \exp\left( -s T_i h_{\mathbf{x}\!\rightarrow\mathbf{y}_0} \lVert \mathbf{x}\rVert^{-\alpha} \right) \right] \right\} \nonumber\\
	&\overset{(a)}{=} \prod_{i\in\left\{1,2\right\}} 
		\exp\left( -\lambda_k \int_{\mathbb{R}^2\!/\!O_{k,i}}
		1 - \frac{1}{1 + s T_i \lVert \mathbf{x}\rVert^{-\alpha}}\, \mathrm{d}\mathbf{x} \right) \nonumber\\
	&\overset{(b)}{=} \prod_{i\in\left\{1,2\right\}} 
		\exp\left( -2\pi\lambda_i 
		\underbrace{
		\int_{\phi_{k,i}\left(x\right)}^\infty
		\frac{r}{1+\left(s T_i\right)^{-1} r^\alpha}\, \mathrm{d}r}_{\text{(c)}}
		 \right),
\end{align}
where the notation $O_{k,i}$ stands for the circle with center at the origin and radius $\phi_{k,i}\left(x\right)$. It is noticed that the interfering BSs in the $i$th tier is farther than $\phi_{k,i}\left(x\right)$ conditioned on the typical UE being associated to the $k$th tier. The step (a) follows from the probability generating functional~(PGFL) of PPP, which converts the multiplication of functions over the point process to an integral, 
and (b) follows from transforming the Cartesian coordination to the polar coordination.  The integral (c) can be evaluated by replacing $r$ with $v^{\frac{1}{2}}$ \cite{gradshteyn2014table}, which gives
\begin{align}\label{eq:app_LT_DL12}
	\mathscr{L}_{I_{\text{DL},k}} \left(s;x\right) 
	&= \exp\left(
		-2\pi\lambda_1 V\left(\phi_{k,1}\left(x\right),\alpha,s T_1\right)
		-2\pi\lambda_2 V\left(\phi_{k,2}\left(x\right),\alpha,s T_2\right) \right),
\end{align}
where
\begin{align}
	V\left(x,\alpha,\beta\right)
	&= \int_x^\infty \frac{r}{1+\beta^{-1}r^\alpha}\,\mathrm{d}r \nonumber\\
	&= \frac{\beta x^{-\alpha+2}}{\alpha-2}{}
		_2F_1\left[1,1-\frac{2}{\alpha};2-\frac{2}{\alpha};-\beta x^{-\alpha}\right].
\end{align}

When the  typical UE is associated to the mmWave SCells in downlink, i.e., $k\in\left\{\text{L,N}\right\}$, the antenna gain of the interfering BS is a discrete random variable $G_\mathrm{b}\left(\theta\right)$ valued at $G_\mathrm{M}$ and $G_\mathrm{m}$, and the value of the interference $I_{\text{DL},k}$ is given by
\begin{align}\label{eq:app_IDLk}
	&I_{\text{DL},k} = 
	\sum_{i\in\left\{\text{L,N}\right\}} \sum_{\mathbf{x}\in\Phi_i \backslash \mathbf{x}_\text{DL}^*} 
	\frac{T_i}{G_\mathrm{M}} 
	G_\mathrm{b}\! \left( \theta_{\mathbf{x} \rightarrow \mathbf{y}_0} \right)
	h_{\mathbf{x}\!\rightarrow\mathbf{y}_0} \lVert \mathbf{x} \rVert^{-\alpha_i}.
\end{align}
Based on \eqref{eq:app_IDLk}, the Laplace transform of $I_{\text{DL},k}$ is formulated as
\begin{align}
	\mathscr{L}_{I_{\text{DL},k}} \left(s;x\right) 
	&= \mathbb{E}_{\Phi,G,h} \left[ \exp\left( -s I_{\text{DL},k}\right) \right] \nonumber\\
	&\overset{}{=} \prod_{i\in\left\{\text{L,N}\right\}} 
		\exp\left( -2\pi\lambda_3 
		\sum_{j\in\left\{\text{M,m}\right\}} p_j 
		\int_{\phi_{k,i}\left(x\right)}^\infty 
		\frac{r P_i\left(r\right) }{1+ \frac{r^\alpha}{s{ T_i }{ {\hat G}_j }}
		  }\, \mathrm{d}r \right) \nonumber\\
	&\overset{}{=} \prod_{i\in\left\{\text{L,N}\right\}} 
		\exp\left( -2\pi\lambda_3 
		\sum_{j\in\left\{\text{M,m}\right\}} p_j 
		W_i \left( \phi_{k,i}, \alpha_i, s T_i \hat{G}_j \right) \right),\label{eq:app_LT_DLLN}
\end{align}
where
\begin{align}
	W_\text{L} \left(x,\alpha_\text{L},\beta\right)
	&= \int_{x}^{\infty}
		\frac{r}{1+\beta^{-1}r^{\alpha_\text{L}}}P_\text{L}\left(r\right)\,\mathrm{d}r, \\
	W_\text{N} \left(x,\alpha_\text{N},\beta\right)
	&= \int_{x}^{\infty}
		\frac{r}{1+\beta^{-1}r^{\alpha_\text{N}}}P_\text{N}\left(r\right)\,\mathrm{d}r.
\end{align}
From \eqref{eq:app_LT_DL12} and \eqref{eq:app_LT_DLLN}, we can derive \eqref{eq:expression_DL_laplace}.

For notational simplicity, we assume that the uplink serving BS $\mathbf{x}_\text{UL}^*$ of the typical UE $\mathbf{y}_0$ is located at the origin. Conditioned on the typical UE $\mathbf{y}_0$ being associated to the sub-6GHz cell in uplink, i.e., $k\in\left\{1,2\right\}$, the interference $I_{\text{UL},k}$ can be written as 
\begin{align}
	I_{\text{UL},k} 
	= \sum\limits_{i\in\left\{1,2\right\}} \sum_{ \mathbf{y}\in\Phi_{u,i} \!\backslash\! \mathbf{y}_0 } 
		T'_i \zeta_\mathbf{y} h_{\mathbf{y}\!\rightarrow\mathbf{x}_\text{UL}^*} \lVert \mathbf{y} \rVert^{-\alpha}.
\end{align}
And the Laplace transform of $I_{\text{UL},k}$ is
\begin{align}
	\mathscr{L}_{I_{\text{UL},k}}\left(s;x\right) 
	&= \mathbb{E}_{\Phi,h,\zeta} \left[ \exp\left( -s I_{\text{UL},k} \right) \right] \nonumber\\
	&\overset{(a)}{=} \prod_{i\in\left\{1,2\right\}} \mathbb{E}_{\Phi_{u,i}} \left\{
		\prod_{ \mathbf{y} \in \Phi_{u,i} \!\backslash\! \mathbf{y}_0 }
		\mathbb{E}_{\zeta_\mathbf{y}} \left[
			\frac{1}{1 \!+\! s T'_i \zeta_\mathbf{y} \lVert \mathbf{y} \rVert^{-\alpha}}  
		\right] \right\} \nonumber\\
	&\overset{(b)}{=} \prod_{i\in\left\{1,2\right\}} \exp\left\{ 
		-2\pi\lambda_i 
		\underbrace{\int_{\varphi_{k,i}\left(x\right)}^\infty 
					\mathbb{E}_{\zeta_\mathbf{y}} \left[
					\frac{r}{1 + \left(s T'_i \zeta_\mathbf{y}\right)^{\!-\!1} \!r^{\alpha}} \right]
				\, \mathrm{d}r}_{\left.J_{k,i}\right|_x} 
		\right\}\label{eq:calculate_LT_UL}
\end{align}
where (a) follows from i.i.d. $h\thicksim\exp(1)$, and (b) follows from the PGFL of PPP. Since $\mathbf{y} \in \Phi_{u,i}\backslash\mathbf{y}_0$, the interfering UE $\mathbf{y}$ is associated to the BS of tier $i$ in uplink, and the inner integral $\left.J_{k,i}\right|_x$ can be calculated as follows.
\begin{align}\label{eq:calculate_J_ki}
	\left.J_{k,i}\right|_x
	&= \int_{\varphi_{k,i}\left(x\right)}^\infty 
		\mathbb{E}_{u_\mathbf{y}} \left[
			\frac{r}{1 + \left(s T'_i u_\mathbf{y}^{\epsilon \alpha}\right)^{\!-\!1} \!r^{\alpha}} \right] 
	\, \mathrm{d}r \nonumber\\
	&= \int_{\varphi_{k,i}\left(x\right)}^\infty 
		\int_0^{\varphi_{k,i}\left(r\right)} 
			\frac{r f_{\mathcal{X}_{\text{UL},i}} \left(u\right)}{1 + \left(s T'_i u^{\epsilon \alpha}\right)^{\!-\!1} \!r^{\alpha}} 
		\, \mathrm{d}u
	\, \mathrm{d}r \nonumber\\
	&\overset{(a)}{=} \int_{0}^{\infty} 
		\int_{ r_0\left(u\right) }^{\infty} 
			\frac{r}{1 + \left(s T'_i u^{\epsilon \alpha}\right)^{\!-\!1} \!r^{\alpha}} 
		\, \mathrm{d}r 
		 f_{\mathcal{X}_{\text{UL},i}} \left(u\right) 
	\, \mathrm{d}u \nonumber\\
	&= \int_{0}^{\infty} 
		V\left( r_0\left(u\right), \alpha, s T'_i u^{\epsilon \alpha} \right) 
		f_{\mathcal{X}_{\text{UL},i}} \left(u\right) 
	\, \mathrm{d}u, 
\end{align}
where $r_0\left(u\right) = \max\left\{ \varphi_{k,i}\left(x\right), \varphi_{i,k}\left(u\right) \right\}$, and (a) follows by interchanging the order of integration. Plugging \eqref{eq:calculate_J_ki} into \eqref{eq:calculate_LT_UL}, we  get the expression of $\mathscr{L}_{I_{\text{UL},k}}\left(s;x\right)$, $k\in\left\{ 1,2\right\}$, as
\begin{align}\label{eq:app_LT_UL12}
	\mathscr{L}_{I_{\text{UL},k}}\left(s;x\right)
	&= \prod_{i\in\left\{ 1,2\right\}} 
		\exp\bigg( -2\pi\lambda_i \int_{0}^{\infty} 
		V\Big( r_0\left(u\right), \alpha, s T'_i u^{\epsilon \alpha} \Big) 
		f_{\mathcal{X}_{\text{UL},i}} \left(u\right) 
	\, \mathrm{d}u \bigg).
\end{align}

Since the UEs associated to the mmWave SCells in uplink transmit with constant power $P_u$, the interference $I_{\text{UL},k}$, $k\in\left\{\text{L,N}\right\}$, can be easily derived from $I_{\text{DL},k}$ in \eqref{eq:app_IDLk} by replacing $P_{\text{DL},k}$ and $\phi_{k,i}\left(x\right)$ with $P_u$ and $\varphi_{k,i}\left(x\right)$, respectively, and it is given by 
\begin{align}\label{eq:app_LT_ULLN}
	\mathscr{L}_{I_{\text{UL},k}}\!\left(\!s;x\!\right)\!
	\!=\! \prod_{i\in\!\left\{\!\text{L,N}\!\right\}\!} 
		\!\exp\!\left(\! -2\pi\lambda_3 
		\!\sum_{j\in\!\left\{\!\text{M,m}\!\right\}\!}\! p_j 
		W_i \!\left(\! \varphi_{k,i}, \alpha_\text{L}, s T'_i \hat{G}_j \!\right)\! \right)\!.
\end{align}
From \eqref{eq:app_LT_UL12} and \eqref{eq:app_LT_ULLN}, we can obtain \eqref{eq:expression_UL_laplace}.

\section{Proof of Theorem~\ref{theorem_sinrcoverage}}\label{appendix:sinrcoverage}
From~\eqref{eq:expression_DL_interference} and~\eqref{eq:expression_DL_laplace}, the SINR coverage probability of tier $k$ is given by
\begin{align}\label{eq:app_Ck}
	\mathcal{C}_{\nu,k} \left(\tau\right)
	&= \mathbb{P}\left( 
		\left.\frac{P_{\nu,k} G_k h \ell_k\!\left(\lVert \mathbf{x}_\nu^* \rVert\right)}{\sigma_k^2 + I_{\nu,k}} > \tau 
		\right| K_\nu = k \right) \nonumber\\
	&= \mathbb{P}\left( 
		\left.h > \frac{\tau\left(\sigma_k^2 + I_{\nu,k}\right)}{S_{\nu,k}\!\left( \lVert\mathbf{x}_\nu^*\rVert \right)} 
		\right| K_\nu = k \right) \nonumber\\
	&\overset{(a)}{=} \mathbb{E}_{\mathbf{x}_\nu^*,I_{\nu,k}}\left[ 
		\left.\exp\left(- \frac{\tau\left(\sigma_k^2 + I_{\nu,k}\right)}{S_{\nu,k}\!\left( \lVert\mathbf{x}_\nu^*\rVert \right)}\right) 
		\right| K_\nu = k \right] \nonumber\\
	&\overset{(b)}{=} \!\int_0^\infty 
		\!\exp\!\left(\!-\frac{\tau\sigma_k^2}{S_{\nu,k}\!\left( x \right)}\!\right)\! 
		\mathscr{L}_{I_{\nu,k}}\!\left(\! \frac{\tau}{S_{\nu,k}\!\left( x \right)\!};x \!\right)\! 
		f_{\mathcal{X}_{\nu,k\!}}\!\left(\!x\!\right)\!
	\, \mathrm{d}x, 
\end{align}
where (a) follows from the complementary CDF of exponential variable $h$, and (b) follows from the definition of Laplace transform $\mathscr{L}_{I_{\nu,k}}\!\left(s;x\right) = \exp\left(-s I_{\nu,k}\right)$. From \eqref{eq:Ck2C} and \eqref{eq:app_Ck}, we can derive \eqref{eq:sinrcov}.


\begin{thebibliography}{99}

\bibitem{Access2013Rappaport}
T.~S. Rappaport, S.~Sun, R.~Mayzus, H.~Zhao, Y.~Azar, K.~Wang, G.~N. Wong,
  J.~K. Schulz, M.~Samimi, and F.~Gutierrez, ``Millimeter wave mobile
  communications for 5{G} cellular: It will work!'' \emph{IEEE Access}, vol.~1,
  pp. 335--349, 2013.

\bibitem{TWC2016Zhang}
H.~Zhang, Y.~Dong, J.~Cheng, M.~J. Hossain, and V.~C.~M. Leung, ``Fronthauling
  for 5{G} {LTE-U} ultra dense cloud small cell networks,'' \emph{IEEE Wireless
  Commun.}, vol.~23, no.~6, pp. 48--53, Dec. 2016.

\bibitem{rappaport2014millimeter}
T.~S. Rappaport, R.~W. Heath~Jr, R.~C. Daniels, and J.~N. Murdock,
  \emph{Millimeter wave wireless communications}.\hskip 1em plus 0.5em minus
  0.4em\relax Pearson Education, 2014.

\bibitem{TC2013Hur}
S.~Hur, T.~Kim, D.~J. Love, J.~V. Krogmeier, T.~A. Thomas, and A.~Ghosh,
  ``Millimeter wave beamforming for wireless backhaul and access in small cell
  networks,'' \emph{IEEE Trans. Commun.}, vol.~61, no.~10, pp. 4391--4403, Oct.
  2013.

\bibitem{JSAC2017Yu_antenna}
X.~Yu, J.~Zhang, M.~Haenggi, and K.~B. Letaief, ``Coverage analysis for
  millimeter wave networks: The impact of directional antenna arrays,''
  \emph{IEEE J. Sel. Areas Commun.}, vol.~35, no.~7, pp. 1498--1512, Jul. 2017.

\bibitem{TWC2015Bai}
T.~Bai and R.~W. Heath, ``Coverage and rate analysis for millimeter-wave
  cellular networks,'' \emph{IEEE Trans. Wireless Commun.}, vol.~14, no.~2, pp.
  1100--1114, Feb. 2015.

\bibitem{TC2017mmTut_Andrews}
J.~G. Andrews, T.~Bai, M.~N. Kulkarni, A.~Alkhateeb, A.~K. Gupta, and R.~W.
  Heath, ``Modeling and analyzing millimeter wave cellular systems,''
  \emph{IEEE Trans. Commun.}, vol.~65, no.~1, pp. 403--430, Jan. 2017.

\bibitem{TWC2016De_Elshaer}
H.~Elshaer, M.~N. Kulkarni, F.~Boccardi, J.~G. Andrews, and M.~Dohler,
  ``Downlink and uplink cell association with traditional macrocells and
  millimeter wave small cells,'' \emph{IEEE Trans. Wireless Commun.}, vol.~15,
  no.~9, pp. 6244--6258, Sep. 2016.

\bibitem{JSAC2012Femtocell_Andrews}
J.~G. Andrews, H.~Claussen, M.~Dohler, S.~Rangan, and M.~C. Reed, ``Femtocells:
  past, present, and future,'' \emph{IEEE J. Sel. Areas Commun.}, vol.~30,
  no.~3, pp. 497--508, Apr. 2012.

\bibitem{CM2013HetNet_Andrews}
J.~G. Andrews, ``Seven ways that {H}et{N}ets are a cellular paradigm shift,''
  \emph{IEEE Commun. Mag.}, vol.~51, no.~3, pp. 136--144, Mar. 2013.

\bibitem{TWC2012HetNet_Dhillon}
H.~S. Dhillon, R.~K. Ganti, F.~Baccelli, and J.~G. Andrews, ``Modeling and
  analysis of {K}-tier downlink heterogeneous cellular networks,'' \emph{IEEE
  J. Sel. Areas Commun.}, vol.~30, no.~3, pp. 550--560, Apr. 2012.

\bibitem{TWC2012HetNet_Jo}
H.~S. Jo, Y.~J. Sang, P.~Xia, and J.~G. Andrews, ``Heterogeneous cellular
  networks with flexible cell association: A comprehensive downlink {SINR}
  analysis,'' \emph{IEEE Trans. Wireless Commun.}, vol.~11, no.~10, pp.
  3484--3495, Oct. 2012.

\bibitem{WC2015HZhang}
H.~Zhang, C.~Jiang, J.~Cheng, and V.~C.~M. Leung, ``Cooperative interference
  mitigation and handover management for heterogeneous cloud small cell
  networks,'' \emph{IEEE Wireless Commun.}, vol.~22, no.~3, pp. 92--99, Jun.
  2015.

\bibitem{CM2014Green}
C.~L. I, C.~Rowell, S.~Han, Z.~Xu, G.~Li, and Z.~Pan, ``Toward green and soft:
  A 5{G} perspective,'' \emph{IEEE Commun. Mag.}, vol.~52, no.~2, pp. 66--73,
  Feb. 2014.

\bibitem{TWC2015Singh_DUDe}
S.~Singh, X.~Zhang, and J.~G. Andrews, ``Joint rate and {SINR} coverage
  analysis for decoupled uplink-downlink biased cell associations in
  {H}et{N}ets,'' \emph{IEEE Trans. Wireless Commun.}, vol.~14, no.~10, pp.
  5360--5373, Oct. 2015.

\bibitem{CM2016DUDe}
F.~Boccardi, J.~Andrews, H.~Elshaer, M.~Dohler, S.~Parkvall, P.~Popovski, and
  S.~Singh, ``Why to decouple the uplink and downlink in cellular networks and
  how to do it,'' \emph{IEEE Commun. Mag.}, vol.~54, no.~3, pp. 110--117, Mar.
  2016.

\bibitem{TWC2017Bacha}
M.~Bacha, Y.~Wu, and B.~Clerckx, ``Downlink and uplink decoupling in two-tier
  heterogeneous networks with multi-antenna base stations,'' \emph{IEEE Trans.
  Wireless Commun.}, vol.~16, no.~5, pp. 2760--2775, May 2017.

\bibitem{CST2017ElSawy_SGtut}
H.~ElSawy, A.~Sultan-Salem, M.~S. Alouini, and M.~Z. Win, ``Modeling and
  analysis of cellular networks using stochastic geometry: A tutorial,''
  \emph{IEEE Communications Surveys Tutorials}, vol.~19, no.~1, pp. 167--203,
  Firstquarter 2017.

\bibitem{conf2012Akoum_mmW}
S.~Akoum, O.~E. Ayach, and R.~W. Heath, ``Coverage and capacity in mm{W}ave
  cellular systems,'' in \emph{2012 Conference Record of the Forty Sixth
  Asilomar Conference on Signals, Systems and Computers (ASILOMAR)}, Nov. 2012,
  pp. 688--692.

\bibitem{JSAC2015Singh_backhaul}
S.~Singh, M.~N. Kulkarni, A.~Ghosh, and J.~G. Andrews, ``Tractable model for
  rate in self-backhauled millimeter wave cellular networks,'' \emph{IEEE J.
  Sel. Areas Commun.}, vol.~33, no.~10, pp. 2196--2211, Oct. 2015.

\bibitem{chiu2013stochastic}
S.~N. Chiu, D.~Stoyan, W.~S. Kendall, and J.~Mecke, \emph{Stochastic geometry
  and its applications}.\hskip 1em plus 0.5em minus 0.4em\relax John Wiley \&
  Sons, 2013.

\bibitem{AWPL2015Colombi}
D.~Colombi, B.~Thors, and C.~Törnevik, ``Implications of {EMF} exposure limits
  on output power levels for 5{G} devices above 6{GHz},'' \emph{IEEE Antennas
  Wirel. Propag. Lett.}, vol.~14, pp. 1247--1249, 2015.

\bibitem{conf2009Mullner_pc}
R.~Mullner, C.~F. Ball, K.~Ivanov, J.~Lienhart, and P.~Hric, ``Contrasting
  open-loop and closed-loop power control performance in {UTRAN} {LTE} uplink
  by {UE} trace analysis,'' in \emph{2009 IEEE International Conference on
  Communications}, Jun. 2009, pp. 1--6.

\bibitem{Bai2014CM_mm}
T.~Bai, A.~Alkhateeb, and R.~W. Heath, ``Coverage and capacity of
  millimeter-wave cellular networks,'' \emph{IEEE Commun. Mag.}, vol.~52,
  no.~9, pp. 70--77, Sep. 2014.

\bibitem{TWC2013Novlan_UL}
T.~D. Novlan, H.~S. Dhillon, and J.~G. Andrews, ``Analytical modeling of uplink
  cellular networks,'' \emph{IEEE Trans. Wireless Commun.}, vol.~12, no.~6, pp.
  2669--2679, Jun. 2013.

\bibitem{CL2017Haenggi_UL}
M.~Haenggi, ``User point processes in cellular networks,'' \emph{IEEE Wireless
  Commun. Lett.}, vol.~6, no.~2, pp. 258--261, Apr. 2017.

\bibitem{TWC2017Ding_UL}
T.~Ding, M.~Ding, G.~Mao, Z.~Lin, D.~López-Pérez, and A.~Y. Zomaya, ``Uplink
  performance analysis of dense cellular networks with {L}o{S} and {NL}o{S}
  transmissions,'' \emph{IEEE Trans. Wireless Commun.}, vol.~16, no.~4, pp.
  2601--2613, Apr. 2017.

\bibitem{CM2017ultradense}
J.~An, K.~Yang, J.~Wu, N.~Ye, S.~Guo, and Z.~Liao, ``Achieve sustainable ultra-dense heterogeneous networks for 5{G},'' 
\emph{IEEE Commun. Mag.}, accepted.

\bibitem{TWC2013Singh_load}
S.~Singh, H.~S. Dhillon, and J.~G. Andrews, ``Offloading in heterogeneous
  networks: modeling, analysis, and design insights,'' \emph{IEEE Trans.
  Wireless Commun.}, vol.~12, no.~5, pp. 2484--2497, May 2013.

\bibitem{JSAC2014Akdeniz}
M.~R. Akdeniz, Y.~Liu, M.~K. Samimi, S.~Sun, S.~Rangan, T.~S. Rappaport, and
  E.~Erkip, ``Millimeter wave channel modeling and cellular capacity
  evaluation,'' \emph{IEEE J. Sel. Areas Commun.}, vol.~32, no.~6, pp.
  1164--1179, Jun. 2014.

\bibitem{gradshteyn2014table}
I.~S. Gradshteyn and I.~M. Ryzhik, \emph{Table of integrals, series, and
  products}.\hskip 1em plus 0.5em minus 0.4em\relax Academic press, 2014.

\end{thebibliography}

\bibliographystyle{IEEEtran}

\end{document}